%% file: main.tex
\documentclass[fleqn,usenatbib]{mnras}

\usepackage{newtxtext,newtxmath}

\usepackage[T1]{fontenc}

\DeclareRobustCommand{\VAN}[3]{#2}
\let\VANthebibliography\thebibliography
\def\thebibliography{\DeclareRobustCommand{\VAN}[3]{##3}\VANthebibliography}

\usepackage{graphicx}	
\usepackage{amsmath}	
\usepackage{booktabs,caption}
\usepackage[flushleft]{threeparttable}
\usepackage{subcaption}
\usepackage{multirow}
\usepackage{dblfloatfix}
\usepackage{pdflscape}
\usepackage{hyperref} 
\usepackage{siunitx}
\usepackage{lipsum}
\usepackage{array}
\usepackage[final]{microtype}
\usepackage{orcidlink}

\newcolumntype{A}{r@{\!}c@{\!}l}
\newcommand{\pmnum}[2]{$#1$ & $\pm$ & $#2$}

\newcommand{\anchornum}[3]{$#1$ & $#2$ & $#3$}
\newcommand{\decnum}[2]{%
  & $\makebox[0pt][r]{$#1$}\mathord{.}\makebox[0pt][l]{$#2$}$ &
}
\newcommand{\intnum}[2]{%
  & $\makebox[0pt][r]{\ensuremath{#1}}\makebox[0pt][l]{\ensuremath{#2}}$ &
}

\title[Low-frequency Galactic emission modeling]{All-sky modeling of Galactic emission at radio and microwave frequencies}

\author[G. A. Hoerning et al.]{Gabriel A. Hoerning$^{\mathrm{\orcidlink{0000-0002-8677-6656}}}$\!,$^{1}$\thanks{E-mail: gabrielamancio.hoerning@postgrad.manchester.ac.uk}
Clive Dickinson$^\mathrm{\orcidlink{0000-0002-0045-442X}}$\!,$^{1}$
Stuart E. Harper$^\mathrm{\orcidlink{0000-0001-7911-5553}}$\!,$^{1}$
Roke Cepeda-Arroita$^\mathrm{\orcidlink{0000-0002-9043-2645}}$\!,$^{2,3}$
\newauthor
Hans K. Eriksen$^\mathrm{\orcidlink{0000-0003-2332-5281}}$\!,$^{4}$
Melis O. Irfan$^{\mathrm{\orcidlink{0000-0003-2021-7357}}\,5}$
J. Patrick Leahy$^\mathrm{\orcidlink{0000-0003-2514-9592}}$\!,$^{1}$
Jamie Leech$^\mathrm{\orcidlink{0000-0002-3615-0727}}$\!,$^{6}$ 
Michael E. Jones$^\mathrm{\orcidlink{0000-0003-3564-6680}}$\!,$^{6}$
\newauthor
Timothy J. Pearson$^\mathrm{\orcidlink{0000-0001-5213-6231}}$\!,$^{7}$
Michael W. Peel$^\mathrm{\orcidlink{0000-0003-3412-2586}}$\!,$^{8}$
Vasundhara Shaw$^\mathrm{\orcidlink{0000-0002-5824-7191}}$\!,$^{1,9}$
Angela C. Taylor$^\mathrm{\orcidlink{0000-0002-3309-9081}}$\!,$^{6}$
\newauthor
Duncan J. Watts$^\mathrm{\orcidlink{0000-0002-5437-6121}}$\!,$^{4}$
Ingunn K. Wehus$^\mathrm{\orcidlink{0000-0003-3821-7275}}$\!,$^{4}$
and Gilles Weymann-Despres$^{\mathrm{\orcidlink{0000-0002-9281-281X}}\,6}$\\
\\
$^{1}$Jodrell Bank Centre for Astrophysics, Department of Physics \& Astronomy, The University of Manchester, Oxford Road, Manchester, M13 9PL, UK\\
$^{2}$Instituto de Astrof\'{i}sica de Canarias, 38200 La Laguna, Tenerife, Canary Islands, Spain\\
$^{3}$Departamento de Astrof\'{i}sica, Universidad de La Laguna (ULL), 38206 La Laguna, Tenerife, Spain\\
$^{4}$Institute of Theoretical Astrophysics, University of Oslo, Blindern, Oslo, Norway\\
$^{5}$Institute of Astronomy, University of Cambridge, Madingly Road, CB3 0HA, UK\\
$^{6}$Department of Physics, University of Oxford, Denys Wilkinson Building, Keble Road, Oxford, OX1 3RH, UK\\
$^{7}$Cahill Centre for Astronomy and Astrophysics, California Institute of Technology, Pasadena, CA 91125, USA\\
$^{8}$Imperial College London, Blackett Lab, Prince Consort Road, London SW7 2AZ, UK\\
$^{9}$FIZ Karlsruhe – Leibniz-Institute for Information Infrastructure, Franklinstr. 11, 10587 Berlin, Germany}

\date{Accepted XXX. Received YYY; in original form ZZZ}

\pubyear{\the\year{}}

\begin{document}
\label{firstpage}
\pagerange{\pageref{firstpage}--\pageref{lastpage}}
\maketitle

\begin{abstract}
We present a new all-sky model of low-frequency diffuse Galactic emission in the regime where synchrotron, free--free, and spinning dust dominate. The model extends the \textit{Planck} 2015 diffuse component-separation analysis by incorporating recent radio and microwave surveys. We fit 35 full- and partial-sky maps at $1^\circ$ resolution, including S-PASS at $2.30$\,GHz, C-BASS at $4.76$\,GHz, and QUIJOTE at $10$--$20$\,GHz, together with reprocessed WMAP and \textit{Planck} LFI data from the Cosmoglobe collaboration and \textit{Planck} HFI channels. Using a Bayesian parametric approach with \texttt{Commander}, we derive spatially varying amplitude and spectral parameter maps for the dominant low-frequency foreground components in total intensity. The main products are a full-sky synchrotron amplitude and spectral-index solution, an all-sky characterization of spinning dust emission with a single-component log-normal spectral model, and a reconstructed all-sky total-intensity map at $4.76$\,GHz tracing diffuse synchrotron emission with reduced systematics relative to Haslam $408$\,MHz. The revised low-frequency anchoring increases the recovered synchrotron amplitude: at $4.76$\,GHz, it is approximately a factor of two higher than the \textit{Planck} 2015 prediction. The model achieves RMS temperature residuals below $10$\,\textmu K over $95\%$ of the sky up to $353$\,GHz, with fractional residuals below $1.5\%$ in the Galactic plane and below $5\%$ across QUIJOTE bands. Residual angular power spectra lie more than two orders of magnitude below the CMB spectrum. These products describe the transition between radio and microwave emission and provide a new reference for foreground modeling and sky-simulation applications.
\end{abstract}

\begin{keywords}
ISM: general -- cosmology: observation -- cosmic microwave background -- diffuse radiation -- Galaxy: general
\end{keywords}



\section{Introduction}\label{sec:int}

Large-scale observations of the radio and microwave sky contain several astrophysical signals originating from both Galactic and extragalactic sources together with the cosmic microwave background (CMB). At low frequencies ($\lesssim10$\,GHz), hereafter referred to as the radio regime, synchrotron emission from the propagation of cosmic rays through the Galactic magnetic field dominates the sky \citep {Strong2011}. Around 10--30\,GHz, thermal free--free emission from ionized gas \citep{Dickinson2003}, and spinning dust emission from the rapid rotation of small grains in the interstellar medium (ISM) \citep{Dickinson2018} become prominent. At higher frequencies ($\gtrsim 100$\,GHz), thermal dust emission from grains in equilibrium with the interstellar radiation field becomes the dominant diffuse Galactic component \citep{planck2013-p06b}. Accurate modeling and separation of these components is crucial for a wide range of astrophysical and cosmological analyses, including characterization of the ISM through cosmic-ray propagation (e.g. \citeauthor{Strong2011} \citeyear{Strong2011}), magnetic turbulence (e.g. \citeauthor{Brandenburg2013} \citeyear{Brandenburg2013}), dust grain physics (e.g. \citeauthor{Draine_book} \citeyear{Draine_book}), the study of Galactic magnetic fields (e.g. \citeauthor{Han2017} \citeyear{Han2017}), and the extraction of the CMB (e.g. \citeauthor{Remazeilles2016} \citeyear{Remazeilles2016}). However, this task remains challenging because, except for free--free emission, the spectral energy distributions (SEDs) of the dominant components are not well constrained a priori, and their relative contributions become comparable in the radio-to-microwave transition region. As a result, degeneracies between synchrotron, free--free, and spinning dust emission remain significant, especially on large angular scales and in the Galactic plane \citep{bennett2003a, leach2008}. These challenges are not only technical, but reflect intrinsic limitations in the available data and frequency coverage.

Several algorithms have been developed to separate these astrophysical components, each making different assumptions about the data and about the nature of the underlying signals. In the recent component separation by the \textit{Planck} Collaboration \citep{planck2016-l04}, the methods that were applied were: \texttt{Commander} (\citeauthor{Eriksen2004} \citeyear{Eriksen2004}, \citeyear{Eriksen2008}, \citeauthor{Seljebotn2019} \citeyear{Seljebotn2019}), the Needlet Internal Linear Combination (\texttt{NILC}; \citeauthor{Basak2012} \citeyear{Basak2012}, \citeyear{Basak2013}), the Spectral Matching Independent Component Analysis (\texttt{SMICA}; \citeauthor{Delabrouille2003} \citeyear{Delabrouille2003}; \citeauthor{cardoso2008} \citeyear{cardoso2008}), and the Spectral Estimation Via Expectation Maximization (\texttt{SEVEM}; \citeauthor{leach2008} \citeyear{leach2008}; \citeauthor{fernandez2012} \citeyear{fernandez2012}) that are capable of recovering the CMB with the level of accuracy required for precision cosmology, whereas approaches like \texttt{Commander}, \texttt{SMICA}, and the Generalized Needlet Internal Linear Combination (\texttt{GNILC}; \citeauthor{Remazeilles2011GNICL} \citeyear{Remazeilles2011GNICL}) could also reconstruct the individual Galactic foregrounds components. As the respective products suggest, these methodologies can be split into two different approaches: ``blind'' methods, such as \texttt{NILC} and \texttt{SEVEM}, that do not assume explicit physical models for the emission mechanisms, but instead use statistical criteria to minimize variance, while the resulting component reconstruction remains subject to implicit constraints from the adopted statistical model and the allowed level of spectral variability; and ``non-blind'' or parametric methods, like \texttt{Commander}, that use physically motivated spectral models for each component and apply Bayesian inference to fit their respective parameters. Some methods, such as \texttt{SMICA} and \texttt{GNILC}, can be called semi-blind, standing between these two cases, which use statistical decompositions of covariance matrices and incorporate prior knowledge. 

The choice between blind and non-blind algorithms is a matter of choosing between robustness to modeling assumptions and physical interpretability. For studies focusing on Galactic foregrounds, where prior knowledge of emission mechanisms is available, parametric methods such as \texttt{Commander} are helpful since they calculate the posteriors for each parameter. These component separation methods are also crucial for algorithms of sky modeling such as the \textit{Planck} Sky Model (PSM; \citeauthor{delabrouille2012} \citeyear{delabrouille2012}) and the Python Sky Model (\texttt{PySM}; \citeauthor{Thorne2017} \citeyear{Thorne2017}, \citeauthor{Zonca2021} \citeyear{Zonca2021}, \citeauthor{Panexp2025} \citeyear{Panexp2025}), that are based on physically motivated descriptions of Galactic emission to generate realistic simulations of the radio and microwave sky. A robust separation of components provides the foundation on which these models are constructed and tested. 

Previous works that aimed to separate Galactic components in the radio and microwave regime have undergone significant evolution over the past two decades. Early analyses based only on WMAP data \citep{Gold2009, Bennett2013} were based on template fitting using external tracers of synchrotron, free--free, and thermal dust emission, such as the 408\,MHz map by \cite{Haslam1982}, H$\alpha$ emission \citep{Finkbeiner2003}, and far-infrared dust templates \citep{finkbeiner1999} respectively. Due to limited frequency coverage ($\approx$20--100\,GHz), these studies could not independently constrain the SED of the low-frequency components. The synchrotron spectral index was typically fixed to a constant value ($\beta_\mathrm{s}\approx-3.1$), and the spinning dust was modeled only implicitly as a dust-correlated residual. Subsequent parametric approaches, including early applications of \texttt{Commander} to the WMAP data \citep{Eriksen2008}, faced similar limitations, as they had to again assume spatially uniform spectral priors. 

The analysis presented in \cite{planck2014-a12}, hereafter referred to as the \textit{Planck} 2015 model, represented a significant step forward by combining \textit{Planck}, WMAP, and Haslam total-intensity data at $1^\circ$ resolution. However, in the absence of complementary low-frequency surveys closer to the CMB frequencies, such as C-BASS (\citealt{Jones2018}, Taylor et al. in prep.), S-PASS \citep{Carretti2019}, and QUIJOTE \citep{rubino2023quijote}, simplified assumptions were still necessary. Two examples are a spatially constant synchrotron spectral index and a phenomenological two-component spinning dust model with fixed SED width. In its latest release, the \textit{Planck} Collaboration \citep{planck2016-l04} adopted a fully self-consistent component separation using only \textit{Planck} data, and no external inputs. They noted, however, that the resulting foreground reconstructions, aside from the CMB, were inferior to those obtained in the \textit{Planck} 2015 model and were thus not publicly released. This outcome underscored the importance of joint analyses combining multiple datasets. More recently, the Cosmoglobe collaboration has provided updated foreground and CMB reconstructions that combine WMAP and \textit{Planck} LFI data \citep{Andersen2023}, as well as a dedicated model of the polarized low-frequency sky \citep{Watts2024}, where both analyses show the power of the data $\lesssim100$\,GHz. With the release of the QUIJOTE maps, the QUIJOTE collaboration has also produced new component separation results in both total intensity \citep{Fernandez2023} and polarization \citep{delaHoz2023,Gonzalez2025}. With the publication of these new low-frequency surveys, it is now possible to overcome several of the limitations that affected earlier analyses, enabling a more accurate and physically motivated description of the low-frequency foregrounds and providing a natural extension of the \textit{Planck} 2015 model.

In this work, we present a new all-sky model of low-frequency diffuse Galactic emission. The model is constructed using a Bayesian parametric component-separation approach implemented with the \texttt{Commander} code, using the following surveys: \textit{Planck} (\citeauthor{planck2016-l02} \citeyear{planck2016-l02}, \citeauthor{planck2014-a09} \citeyear{planck2014-a09}, \citeauthor{Watts2023} \citeyear{Watts2023}), WMAP (\citeauthor{Bennett2013} \citeyear{Bennett2013}, \citeauthor{Watts2023} \citeyear{Watts2023}), C-BASS, S-PASS, and QUIJOTE. \texttt{Commander} is the appropriate choice for our goal, as it enables us to use physical priors and model the SED of each component. This paper focuses on total intensity (Stokes~\textit{I}), with polarization left for future work. Our analysis follows a philosophy similar to that of the \textit{Planck} 2015 model, in which the derived parametric model serves as our benchmark for comparing our products. Each survey contributes complementary information that helps to break degeneracies among the components, improving the reconstruction. Unlike the studies mentioned above, we do not use the Haslam 408\,MHz map as a reference for synchrotron emission. By leveraging contemporary, well-calibrated datasets, we aim to achieve a more physically motivated model of the radio and microwave sky. This approach also allows us to reassess long-standing assumptions in Galactic emission modeling, such as the use of the Haslam map as a synchrotron tracer, and to explore how recent surveys reshape our understanding of the low-frequency sky.  

This paper is organized as follows. Section~\ref{sec:data} describes the datasets used and the associated processing steps. Section~\ref{sec:skyModel} summarizes the \texttt{Commander} code and the physical motivation for our parametric sky model. Section~\ref{sec:results} presents the resulting component reconstructions and compares them with the \textit{Planck} 2015 model. In Section~\ref{sec:discussion}, we assess the reliability and limitations of the model by comparing with existing results in the literature. Our main conclusions are summarized in Section~\ref{sec:conclusions}.

\section{DATA SELECTION AND PROCESSING} \label{sec:data}
The data set is assembled to anchor each foreground component close to the frequency range where it dominates, reducing the need for large extrapolations between widely separated radio and microwave regimes. A summary of the data sets is provided in Table~\ref{tab:data}. Surveys near a few gigahertz constrain synchrotron emission, intermediate-frequency data inform the transition between synchrotron, free--free, and spinning dust, and far-infrared channels anchor thermal dust. All maps are convolved to a common angular resolution of $1^\circ$. Per-pixel uncertainties are taken from the survey white-noise RMS maps, and the typical noise level of each channel at this resolution is reported in the Noise RMS column of Table~\ref{tab:data}, computed as the mean value of the corresponding RMS map after smoothing to $1^\circ$. The analysis in this work is done at $\mathrm{N_{\textrm{SIDE}}}=256$, corresponding to $7^\prime \times 7^\prime$ pixels. For each map, the smoothing is performed in harmonic space by applying the ratio of the target and native instrumental-plus-pixel transfer functions to the spherical-harmonic coefficients,
\begin{equation}
    a_{\ell m}^{\rm out}
    =
    a_{\ell m}^{\rm in}
    \left(
    \frac{B_{\ell}^{1^\circ}}{B_{\ell}^{\rm native}}
    \right)
    \left(
    \frac{P_{\ell}^{\rm out}}{P_{\ell}^{\rm native}}
    \right),
\end{equation}
where \(P_{\ell}^{\rm out}\equiv P_{\ell}^{\rm N_{\rm SIDE}=256}\) and \(P_{\ell}^{\rm native}\) are the HEALPix pixel-window functions at the output and native map resolutions, respectively.

\input{tables/data_sets}

\subsection{S-PASS}\label{sssec:spass}
The S-band Polarization All Sky Survey (S-PASS) is a southern-sky survey conducted with the Parkes telescope at 2.30\,GHz \citep{Carretti2019}, with an angular resolution of 8.9 arcmin and coverage at declinations $\delta < -1^\circ$. At this frequency, synchrotron emission dominates the diffuse sky brightness, making S-PASS a key anchor for the low-frequency synchrotron component in the southern hemisphere. The survey has an absolute calibration uncertainty of $5\%$. S-PASS and C-BASS overlap over roughly $16^\circ$ in declination. In this region, comparison of the two surveys in internal analyses reveals residual structures inconsistent with white noise, likely associated with ground pickup and scan-synchronous effects typical of ground-based large-area radio surveys. We exclude S-PASS data in the overlap region and retain only the C-BASS data, owing to its closer frequency proximity to WMAP and \textit{Planck} and the more extensive characterization of its systematics. In the overlap region, including S-PASS would add limited independent constraining power while increasing the risk of propagating residual systematic differences into the fit. As no detailed beam model is publicly available, we adopted the native S-PASS beam as a Gaussian with FWHM 8.9 arcmin before convolution to the common $1^\circ$ resolution. The S-PASS collaboration provides RMS maps for polarization only, so to construct an intensity uncertainty estimate, we use these RMS maps together with an additional regularization noise of $600$~\si{\micro\kelvin}, which yields residuals comparable to those of C-BASS in overlapping regions and avoids overweighting S-PASS in the fit.

\subsection{C-BASS}\label{ssec:cbass}
The C-Band All-Sky Survey (C-BASS) provides large-area intensity maps at 4.76\,GHz with an angular resolution of 47 arcmin \citep{Jones2018}. In this work we use the northern survey (Taylor et al. in prep.), covering $\delta > -14.\!\!^\circ6$. At this frequency, synchrotron emission dominates over free--free and spinning dust at most latitudes, placing C-BASS at a critical frequency for component-separation analysis. It provides one of the highest-frequency large-area tracers of synchrotron emission for which contamination from free--free and spinning dust remains comparatively limited. C-BASS serves as the primary anchor for the synchrotron component in the northern sky, replacing Haslam $408$\,MHz map and reducing sensitivity to low-frequency spectral curvature. In combination with S-PASS in the southern hemisphere, it provides nearly full-sky coverage at frequencies substantially higher than classical low-frequency radio surveys, improving the separation between synchrotron, free--free, and spinning dust emission in the 5--20\,GHz range \citep{Jew2019}. As a single-dish survey, C-BASS is affected by instrumental effects typical of large-area ground-based radio maps, including ground pickup and residual scan-synchronous structures. These effects are well characterized in the survey data release and are constrained to the few-per-cent level in amplitude over the relevant multipole ranges, corresponding to sub-mK fluctuations ($\sim0.3$--$1$\,mK for typical $I$-band sky levels) on degree scales. The survey benefits from a stable instrumental design and a detailed and observationally-confirmed beam model (Leech et al. in prep), with an absolute calibration uncertainty of $2.5\%$. The native beam is convolved to the common $1^\circ$ resolution used in the fit.

\subsection{QUIJOTE}\label{sssec:quijote}
The Q-U-I JOint TEnerife (QUIJOTE) experiment is a multi-frequency microwave survey conducted at the Teide Observatory, covering 10--40\,GHz with angular resolutions ranging from $39$ arcmin to $56$ arcmin (\citeauthor{Rubino2012} \citeyear{Rubino2012}, \citeyear{rubino2023quijote}). In this work, we use total-intensity data from the Multi-Frequency Instrument (MFI) at $11.1$, $12.9$, $16.8$, and $18.8$\,GHz bands. These frequencies probe the low-frequency side of spinning dust emission and, in combination with higher-frequency WMAP and \textit{Planck} channels, help constrain its peak frequency and spectral width. The QUIJOTE beams are well characterized, and the native beams are convolved to the common $1^\circ$ resolution. The calibration uncertainty of the MFI channels is $5\%$.

\begin{figure}
    \centering
    \includegraphics[width=\linewidth]{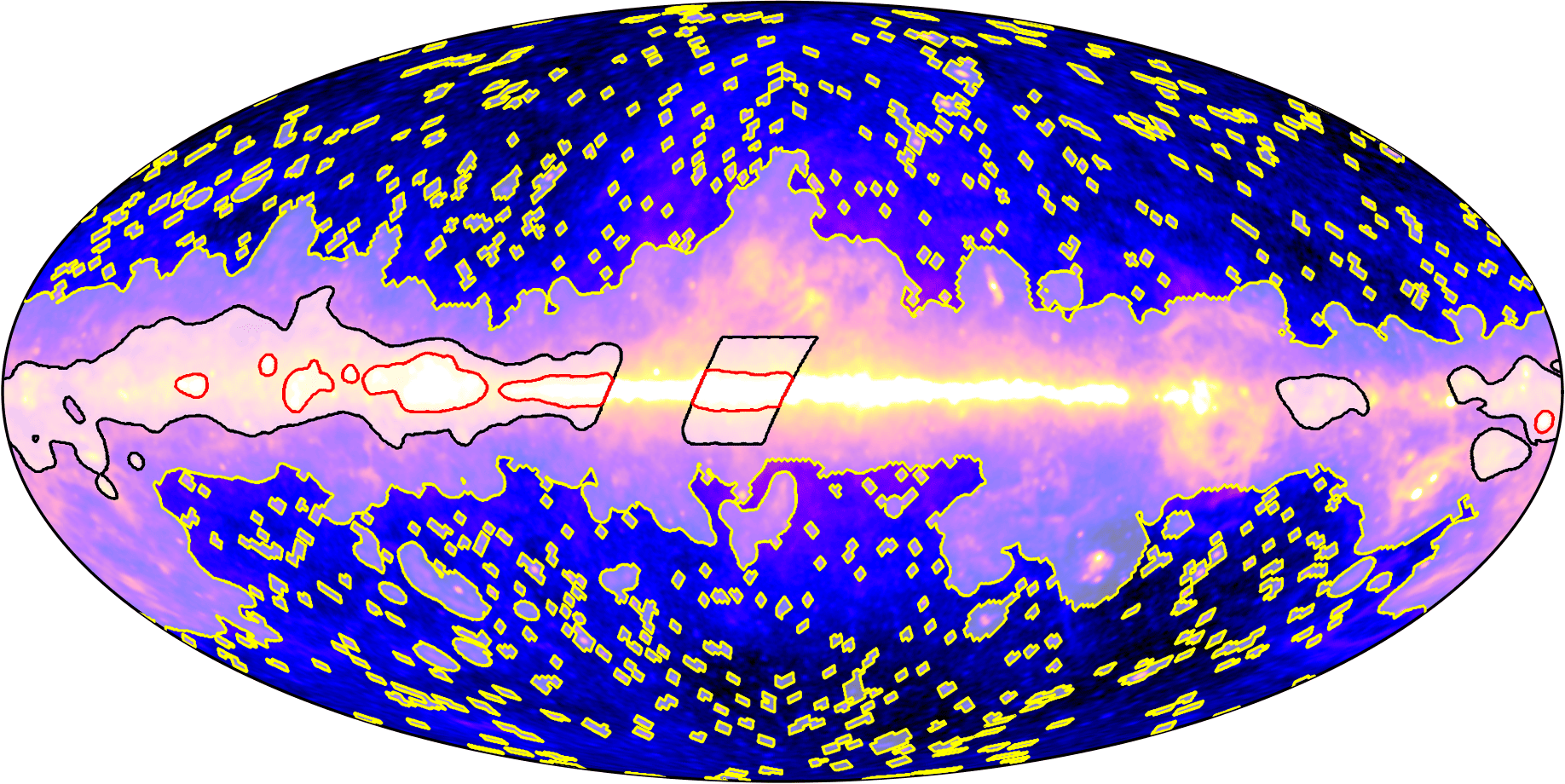}
    \caption{WMAP K-band ($22.8$\,GHz) map with the CMB signal subtracted, overlaid with the masks used for free--free emission modeling and QUIJOTE data. \textit{Black contours:} Galactic plane mask for QUIJOTE 11 and 13\,GHz. \textit{Red contours:} Galactic plane mask for QUIJOTE 17 and 19\,GHz. \textit{Yellow contours:} FF47 mask. Highlighted regions are where pixel fitting for free--free emission is performed, otherwise, the H$\alpha$ map is adopted with the respective \si{\micro\kelvin}/R ratio.}
    \label{fig:masks}
\end{figure}

Residual radio-frequency interference (RFI) increases the effective noise level of the MFI maps, so that the weak high-latitude signal is detected with insufficient signal-to-noise ratio (S/N) for robust component separation. In addition, the processing applied to mitigate RFI, also filters out large angular-scale modes. In principle, this effect should be treated through a forward model of the instrumental transfer function, but such a treatment is beyond the scope of the present analysis. We therefore restrict the use of QUIJOTE to bright Galactic-plane regions, where the sky signal is sufficiently strong to provide useful constraints. For the $11.1$ and $12.9$\,GHz channels, we apply a temperature threshold of $2$\,mK directly on the public maps, while for the $16.8$ and $18.8$\,GHz channels we adopt a threshold of $4$\,mK. The resulting masks are shown in Fig.~\ref{fig:masks}.

Internal \texttt{Commander} tests show that using only the thermal RMS maps for these channels leads to significant tension with the rest of the data set, indicating that additional uncertainty beyond the provided RMS is required for an internally consistent multi-survey fit. To account for this, we increase the per-pixel uncertainty by adding in quadrature a term corresponding to $5\%$ of the local map intensity, thereby down-weighting QUIJOTE in the joint fit. This additional variance does not correct the filtering or restore the lost large-scale information, and residual bias may remain in the recovered amplitudes and spectral parameters. Despite this conservative treatment, the QUIJOTE channels retain sufficient constraining power to improve the determination of the synchrotron, free--free, and spinning dust spectral parameters.

\subsection{WMAP}\label{sssec:wmap}

We use the 9-year Wilkinson Microwave Anisotropy Probe (WMAP) temperature maps \citep{Bennett2013}, as reprocessed by the Cosmoglobe collaboration \citep{Watts2023}. WMAP provides full-sky coverage from 23 to 94\,GHz, bridging the transition between radio and microwave emission and contributing to the separation of synchrotron, free--free, and spinning dust components. The Cosmoglobe reprocessing improves the treatment of large-scale systematics and calibration relative to the original WMAP releases, particularly on the largest angular scales, which are critical for constraining diffuse Galactic emission. All WMAP channels are convolved to the common $1^\circ$ resolution used in this analysis, using the native beam transfer functions. The absolute calibration uncertainty is $1\%$. Residual beam asymmetries present in the Cosmoglobe maps introduce small-scale artifacts around bright compact sources. To mitigate these artifacts we use a hybrid large-scale/small-scale combination between the WMAP Cosmoglobe and DR5 maps described in Appendix~\ref{app:wmap_hybrid}. Outside bright compact sources, these differences have negligible impact on the diffuse emission targeted in this work.

\subsection{\textit{Planck}}\label{sssec:planck}

We use maps from the \textit{Planck} Low Frequency Instrument (LFI) reprocessed by the Cosmoglobe collaboration \citep{Watts2023}, together with High Frequency Instrument (HFI) detector-set maps from the \textit{Planck} 2015 release \citep{planck2014-a08}. The LFI channels ($28.4$–$70.4$\,GHz) bridge the transition between synchrotron-dominated emission and the regime where free--free and spinning dust become important, while the HFI channels (100–857\,GHz) anchor the thermal dust component. The Cosmoglobe reprocessing improves the treatment of large-scale systematics and calibration in the LFI maps, which is critical for modeling diffuse Galactic emission. We adopt the \textit{Planck} 2015 beam profiles when smoothing the LFI maps to the common $1^\circ$ resolution, as use of the 2018 beams was found to introduce non-astrophysical artifacts in residual maps aligned with the HEALPix projection near the Galactic poles. The HFI 2015 detector-set maps provide access to single-bolometer data, facilitating the separation of CO and HCN line emission in the sky model (Section~\ref{subsec:skymodel}), and are the same data sets used in the \textit{Planck} 2015 model. The calibration uncertainty is $1\%$ for all \textit{Planck} channels, except for 353\,GHz ($1.3$\%), 545\,GHz ($6.0$\%), and 857\,GHz ($6.4$\%).

\section{SKY MODEL AND INFERENCE} \label{sec:skyModel}
In this section, we define the parametric sky model adopted in \texttt{Commander}, including the diffuse components, their SEDs, and the physical motivation for each term. The main modifications relative to previous implementations concern the low-frequency foregrounds, while the treatment of molecular line emission, thermal dust, and instrumental parameters follows the approach adopted in the \textit{Planck} 2015 model.

\subsection{COMMANDER} \label{subsec:commander}
We perform the component separation using the Bayesian parametric framework \texttt{Commander} \citep{Eriksen2004,Eriksen2008,planck2014-a12}, which fits an explicit physical model of the sky to multi-frequency observations in pixel space. For each frequency channel $\nu$, the data are modeled as
\begin{equation}
\mathbf{d}_\nu = g_{\nu} \sum_i F^i_\nu(\boldsymbol{\beta}_i, \Delta_\nu)\, \mathbf{a}_i + T_\nu\, \mathbf{m}_\nu + \mathbf{n}_\nu ,
\end{equation}
where $\mathbf{a}_i$ are component amplitude maps at reference frequencies, $\boldsymbol{\beta}_i$ are spectral parameters, $g_{\nu}$ is a multiplicative calibration factor for frequency $\nu$, and $F^i_\nu$ encodes the frequency dependence, including bandpass and unit-conversion effects. The parameter $\Delta_\nu$ denotes effective bandpass corrections. The term $T_\nu\,\mathbf{m}_\nu$ represents additive template corrections, where $\mathbf{m}_\nu$ includes the monopole, the three dipole components, and, where relevant, the CMB kinematic-quadrupole template, while $T_\nu$ denotes the corresponding fitted template amplitudes. Finally, $\mathbf{n}_\nu$ is Gaussian noise with known covariance. The parameter estimation proceeds by sampling the posterior distribution under Gaussian noise assumptions with informative priors that reduce degeneracies between components, including positivity constraints on amplitudes (except the CMB), Gaussian priors on spectral parameters, and fixed monopole, dipole, and calibration constraints for selected channels. We adopt the pixel-space implementation of \texttt{Commander} rather than the harmonic-space \texttt{Commander3} \citep{Seljebotn2019,Galloway2023}. Our goal is to recover spatially resolved spectral parameters and construct spectral index maps from the posterior distributions. Harmonic-space component separation mixes emission transverse to the line-of-sight through the spherical-harmonic decomposition, effectively averaging spectral variations and limiting the recovery of local spectral information. The pixel-based approach allows direct estimation of spatially localized spectral parameters, which is essential for the scientific objectives of this work. The analysis is based on a Gibbs chain of 40\,000 samples, from which the first 4\,000 are discarded as burn-in and all posterior summaries presented in this work are computed from the remaining samples.

\subsection{SKY MODEL}\label{subsec:skymodel}
We now specify the spectral models adopted for each diffuse sky component and the physical processes they represent. The corresponding SEDs and prior values are summarized in Table~\ref{tab:sed}.\\

\input{tables/sed}

\noindent\textit{CMB} --- The CMB SED is described by a blackbody spectrum with a single parameter, the CMB temperature $T_{\rm cmb}$. In practice, the monopole is removed from all maps and \texttt{Commander} models only the CMB anisotropies, whose frequency dependence is given by the derivative of the blackbody spectrum evaluated at $T_{\rm cmb}$. We adopt $T_{\rm cmb}=2.7255\,$K \citep{Fixsen2009} and fix this value in the analysis, as its uncertainty is negligible compared to other sources of error. The Cosmoglobe reprocessing includes a correction for the relativistic kinematic quadrupole that can affect low-frequency \textit{Planck} LFI and WMAP channels \citep{Notari2015}. Nevertheless, we find that a small residual remains in the LFI 70\,GHz map. To account for this, we include a kinematic-quadrupole template evaluated at the central frequency of this channel with a free amplitude determined through maximum-likelihood fitting. This amplitude is not sampled in the posterior analysis. The corresponding effect is negligible for the HFI channels.\\

\noindent\textit{Synchrotron} --- Diffuse synchrotron emission dominates the sky at gigahertz frequencies and arises from relativistic electrons spiraling in the Galactic magnetic field \citep{Strong2011}. Above a few gigahertz, the synchrotron spectrum is well described by a power-law with spectral index $\beta_{\rm s}\approx-3.1$ \citep{Reich1988,kogut2007,Harper2022}, while significant spectral flattening occurs at lower frequencies \citep{deOliveira-Costa2008,Kogut2012}. Since our low-frequency anchors (S-PASS at 2.30\,GHz and C-BASS at 4.76\,GHz) lie in the range where a simple power-law approximation is valid, we model the synchrotron SED with spatially varying amplitude $A_{\rm s}$ and spectral index $\beta_{\rm s}$. In contrast to earlier analyses, we do not include the Haslam 408\,MHz map, as connecting the sub-gigahertz regime to gigahertz frequencies would require an additional curvature parameter and the map contains known large-scale systematics that can dominate fits due to its high S/N \citep{Remazeilles2015,Wilensky2025,Nasirudin2025}. By restricting the model to frequencies above 2\,GHz, we avoid the regime of strong spectral curvature and retain a physically motivated power-law description. To control degeneracies with free--free and spinning dust emission, we adopt a Gaussian prior $\beta_{\rm s}\sim N(-3.1,\,0.3)$, which provides sufficient flexibility while preventing unphysical solutions.\\

\noindent\textit{Free--free} --- Free--free (bremsstrahlung) emission arises from electron–ion collisions in the ionized ISM and has a well-defined spectral form governed by plasma physics \citep{Dickinson2003,Draine_book}. We adopt the same two-parameter description as in the \textit{Planck} 2015 model, with emission measure (EM) and electron temperature $T_{\mathrm{e}}$, but use an H$\alpha$ template \citep{Dickinson2003} as an external tracer to reduce degeneracies with synchrotron and spinning dust at high Galactic latitudes, where free--free emission is weak. The H$\alpha$-derived free--free brightness temperature is modeled with a fixed spectral index $\beta_{\rm ff}=-2.12$ and an amplitude of $195\,$\si{\micro\kelvin}\,R$^{-1}$ at 4.76\,GHz (equivalent to $7.04\,$\si{\micro\kelvin}\,R$^{-1}$ at 22.8\,GHz; \citealt{Harper2022}). Because H$\alpha$ is strongly affected by dust extinction near the Galactic plane, we construct an extinction mask using the thermal-dust optical depth at 353\,GHz, excluding pixels with $\tau_{353}>10^{-5}$ after smoothing to 3$^\circ$. Using the dust-to-reddening conversion from \citet{planck2013-p06b}, this threshold corresponds approximately to $E(B-V)\simeq0.15$, and therefore identifies regions where dust attenuation makes the H$\alpha$ template increasingly unreliable as a free--free tracer. We combine this mask with the PM61 mask from the \textit{Planck} 2015 model to define the FF47 mask, shown in Fig.~\ref{fig:masks}. Outside the masked regions, the scaled H$\alpha$ template is used directly, while inside the mask free--free is determined through pixel-based fitting.

Internal tests showed that freely sampling the H$\alpha$ amplitude at 22.8\,GHz led to its suppression in the MCMC solution, with the missing signal absorbed by synchrotron or spinning dust due to their greater spectral flexibility. To avoid this leakage, we fix the H$\alpha$ amplitude to $7.04\,$\textmu\si{\micro\kelvin}\,R$^{-1}$ at 22.8\,GHz. At high Galactic latitudes the free--free signal is only of order a few \textmu K at this frequency, and does not drive the fit. In parts of the Galactic plane, degeneracies occasionally drove the fitted free--free amplitudes below the H$\alpha$-scaled expectation. To prevent unlikely solutions, we apply a constraint similar to \citet{Gold2009}, requiring $T_{\mathrm{b,ff}}(\hat{\boldsymbol n}) > 0.5\cdot\,7.04\,{\rm H}\alpha(\hat{\boldsymbol n})$, and otherwise reverting to the scaled template. Finally, the impact of varying $T_{\mathrm{e}}$ on the recovered foreground parameters is negligible. We fix $T_{\mathrm{e}}$ to the values provided by the \textit{Planck} 2015 model.\\

\noindent\textit{Spinning Dust} --- Dust grains emit microwave radiation through rotational electric-dipole emission when they possess a permanent dipole moment. In this analysis, anomalous microwave emission is modeled exclusively as spinning dust emission. Other proposed mechanisms, such as magnetic dipole emission from magnetic grains \citep{Draine1999,Draine2013}, are not included. We describe this spinning dust component using a phenomenological log-normal SED, which provides sufficient flexibility to reproduce the observed spectra given the current frequency coverage and calibration limitations while remaining constrained by the data \citep{Bonaldi2007,Stevenson2014_lognormal,Dickinson2018,Cepeda-Arroita2021,Poidevin2023,Cepeda-Arroita2025}. The model has three parameters: the amplitude \(A_{\mathrm{sd}}\), the peak frequency \(\nu_{\mathrm{p}}\) (defined in flux-density units), and the spectral width \(w\) (Table~\ref{tab:sed}). Although \texttt{Commander} fits the SED in Rayleigh--Jeans (RJ) temperature units, we define \(\nu_{\mathrm{p}}\) as the peak frequency of the corresponding flux-density spectrum. We convert the spinning dust SED to flux-density units by multiplying by \((\nu/\nu_{0,\mathrm{sd}})^2\), and estimate \(\nu_{\mathrm{p}}\) from the maximum of the converted spectrum. This approach differs from the \textit{Planck} 2015 model, which employed a two-component SED based on \texttt{SpDust2} templates \citep{Ali-Hamoud2009,Ali-Haimoud2010,Silsbee2011}. Those spectral templates assume fixed spectral widths and specific interstellar environments, whereas the log-normal form allows the spectral width to vary spatially and better captures the broader spectra observed in real data \citep{Cepeda-Arroita2025}. The peak frequency depends on the local grain properties and environment and is typically observed between 20–30\,GHz, with higher values in some regions \citep{planck2013-XV,Fernandez2023,Cepeda-Arroita2025}. We adopt Gaussian priors \( \nu_{\mathrm{p}} \sim N(25,\,5\,\mathrm{GHz}) \) and \( w \sim N(0.6,\,0.1) \). These priors stabilize the solution in low S/N regions and help mitigate degeneracies among low-frequency foreground components without forcing the fit in well-constrained areas. \\

\noindent\textit{Other components} --- The modeling of zodiacal emission, thermal dust, thermal Sunyaev–Zeldovich (SZ), and molecular line emission follows the strategy adopted in the \textit{Planck} 2015 component-separation analysis. The constraints on these components arise primarily from the \textit{Planck} HFI 2015 frequency channels, which are also used here, and we retain the same parametric forms and priors as in that work. Residual Zodiacal emission in WMAP and LFI channels is corrected through template subtraction using multiplicative factors from Table~2 of the \textit{Planck} 2015 model, based on the 100\,GHz HFI Zodiacal light template \citep{planck2013-pip88}.

Thermal dust emission is described by a single-component modified blackbody spectrum with three free parameters per pixel: amplitude \(A_{\mathrm{d}}\), emissivity index \(\beta_{\mathrm{d}}\), and dust temperature \(T_{\mathrm{d}}\) \citep{planck2013-p06}. This model captures the large-scale Galactic dust SED over the frequency range considered (up to 857\,GHz) and is consistent with the approach adopted in \textit{Planck} 2015. More complex dust models (e.g. \citealt{Gjerlow2026}), can also provide improved fits within the \textit{Planck} frequency range, but are not required for the present analysis, which does not focus on the detailed modeling of the HFI channels.

The thermal SZ effect arises from inverse-Compton scattering of CMB photons by hot electrons in galaxy clusters \citep{Sunyaev&Zeldovich1970}. It is modeled using a single amplitude parameter, the Compton-\(y\) parameter \(y_{\mathrm{sz}}\), with the frequency dependence fixed by the SZ spectral distortion (Table~\ref{tab:sed}). Because the SZ signal is weak over most of the sky and susceptible to leakage from other components, we restrict SZ fitting to the regions around the Coma and Virgo clusters following \citet{planck2014-a12}, thereby limiting contamination of the diffuse foreground solution.

Molecular line emission from CO is modeled with separate components for the \(J=1\!\rightarrow\!0\), \(J=2\!\rightarrow\!1\), and \(J=3\!\rightarrow\!2\) transitions, which primarily affect the \textit{Planck} 100, 217, and 353\,GHz channels, respectively \citep{planck2013-p03a}. We also include the ``94/100\,GHz'' line-emission component introduced in the \textit{Planck} 2015 model, which captures molecular-line emission detected through differences between the WMAP W-band and HFI 100\,GHz bandpasses. This component includes a significant contribution from HCN \(J=1\!\rightarrow\!0\) at 88.6\,GHz, along with contributions from other molecular lines and CO leakage \citep{Takekawa2014}. Each line component is described by an amplitude map \(A_i\), normalized to a reference detector. Frequency scaling is implemented using spatially constant line ratios \(h_{ij}\) and unit-conversion factors \(F(\nu)\), accounting for bandpass and unit-conversion effects. Since the relevant channels and bandpasses match those in the \textit{Planck} 2015 model, we fix the \(h_{ij}\) parameters to their 2015 values and sample only the amplitudes \(A_i\).\\

\noindent\textit{Monopoles and dipoles} --- The lowest multipoles of the sky maps, corresponding to the monopole (\(\ell=0\)) and dipole (\(\ell=1\)) modes, are fundamentally degenerate with instrumental offsets. Differential microwave observations constrain spatial variations of the sky brightness but do not directly measure the absolute sky temperature. Physically, the sky monopole is dominated by the CMB blackbody temperature of 2.7255\,K, followed at high frequencies by the mean level of the Cosmic Infrared Background (CIB) \citep{planck2013-pip56}, while the dominant dipole arises from the CMB kinematic dipole, measured to be 3.3655\,mK and 3.3640\,mK by LFI and HFI respectively \citep{planck2014-a01}. However, these absolute contributions are not preserved in most microwave maps. As a consequence, only angular structure at multipoles \(\ell \geq 2\) is robustly determined from the data alone, and monopole and dipole terms must be treated as offsets within the component-separation.

This limitation is reinforced by the calibration scheme of the data sets used here. WMAP and \textit{Planck} maps are calibrated relative to the CMB dipole and processed to remove large-scale offsets, such that the mean temperature at high Galactic latitudes is close to zero by construction. As a result, these maps do not preserve the absolute sky zero level, but only relative temperature fluctuations about an arbitrary reference. Although C-BASS provides a calibration tied to ARCADE\,2 (Leahy et al. in prep.), this cannot be adopted directly in a joint analysis unless all channels share a consistent absolute reference, which is not the case. The present work operates in a relative framework, solving for monopole and dipole offsets instead than attempting to determine the true zero level of the sky. Following the strategy of the \textit{Planck} 2015 model, we anchor a subset of channels and solve for residual monopole and dipole offsets in the remaining maps using the PM61 mask. This procedure yields a self-consistent large-scale solution for all astrophysical components while remaining insensitive to the absolute sky temperature.

The calibration of the low-frequency radio surveys is important, since residual offsets and scale differences directly affect the amplitudes and spectral indices. In this work, C-BASS sets the synchrotron amplitude scale in the northern hemisphere, while S-PASS provides the corresponding reference in the southern hemisphere through alignment in their overlap region. QUIJOTE offsets are determined within the Galactic-plane region used in the fit to ensure internal consistency of the spectral solution. Cosmoglobe-reprocessed WMAP and LFI maps retain the \textit{Planck} 2015 model for monopole and dipole normalization scheme with minor adjustments for the reprocessed maps. Since C-BASS and S-PASS define the reference amplitude scale for the synchrotron component in the northern and southern hemispheres, respectively, their 5\% absolute calibration uncertainties propagate as multiplicative systematic uncertainties into the recovered synchrotron brightness and spectral index maps. The detailed procedures used to determine the offsets of the low-frequency surveys are described in Appendix~\ref{app:md}. A fully self-consistent determination of the absolute sky zero level lies beyond the scope of this work and will require future absolutely calibrated experiments, such as ARCADE\,2 \citep{Fixsen2011}, L-BASS \citep{Zerafa2025}, and the Tenerife Microwave Spectrometer \citep{TMS2020}.\\

\noindent\textit{Instrumental effects} --- In addition to the astrophysical parameters, the model formally includes instrumental degrees of freedom following the scheme adopted in the \textit{Planck} 2015 model. These consist of multiplicative gain factors \(g_\nu\) and effective bandpass shifts \(\Delta_\nu\), which modify the mixing matrix and absorb residual inter-channel calibration and spectral-response uncertainties. Such parameters are important because calibration mismatches project directly onto the inferred spectral behavior of the components.

In the present work, we do not sample gain parameters for C-BASS, S-PASS, or QUIJOTE. For C-BASS and S-PASS, the gain is strongly degenerate with the synchrotron amplitude, as these channels define the low-frequency anchor of that component and set its overall amplitude scale. Allowing $g_\nu$ to vary would primarily redistribute power between instrumental gain and astrophysical amplitude without introducing additional constraining information. For QUIJOTE, the limited sky fraction used in the analysis leads to a similar degeneracy with local foreground amplitudes, preventing an independent determination of the gain. The gain parameters for these channels are held fixed. Likewise, we do not fit bandpass-shift parameters for C-BASS, S-PASS, or QUIJOTE. Instead, the finite bandpasses of these experiments ($\approx20$\% for C-BASS, $\approx7$\% for S-PASS, and $\approx15$\% for QUIJOTE) are accounted for through color corrections. These corrections are implemented using second-order polynomial expansions of the bandpass response (e.g. \citealt{Peel2022}), which are incorporated directly into the model prediction. For typical synchrotron spectral indices ($\beta \sim -3.1$), neglecting bandpass integration would bias the recovered spectral index by of order $\sim2$\% for C-BASS and below the percent level for S-PASS and QUIJOTE. In this way, bandpass effects are treated consistently while avoiding additional degenerate instrumental degrees of freedom. For the remaining channels, the treatment of instrumental parameters and priors follows the \textit{Planck} 2015 model.

\section{Results}\label{sec:results}
In this section, we present the main results obtained for the CMB and the low-frequency Galactic foregrounds. The discussion focuses on the improvements relative to the \textit{Planck} 2015 model, which primarily arise from the revised treatment of low-frequency data and foreground components. High-frequency foregrounds ($\gtrsim100$\,GHz), including thermal dust emission and molecular line emission, are modeled following the same approach as in the previous analysis and show no significant differences and they are not discussed further here. We begin by presenting the results for the CMB, synchrotron, free--free, and spinning dust emission, followed by an assessment of the goodness-of-fit. The section concludes with one of the main products of this work: a reconstructed all-sky map at 4.76\,GHz. The posterior constraints on instrumental parameters, monopoles, and dipoles are summarized in Appendix~\ref{app:md} (Table~\ref{tab:md_inst}).

\begin{figure*}
	\includegraphics[width=\textwidth]{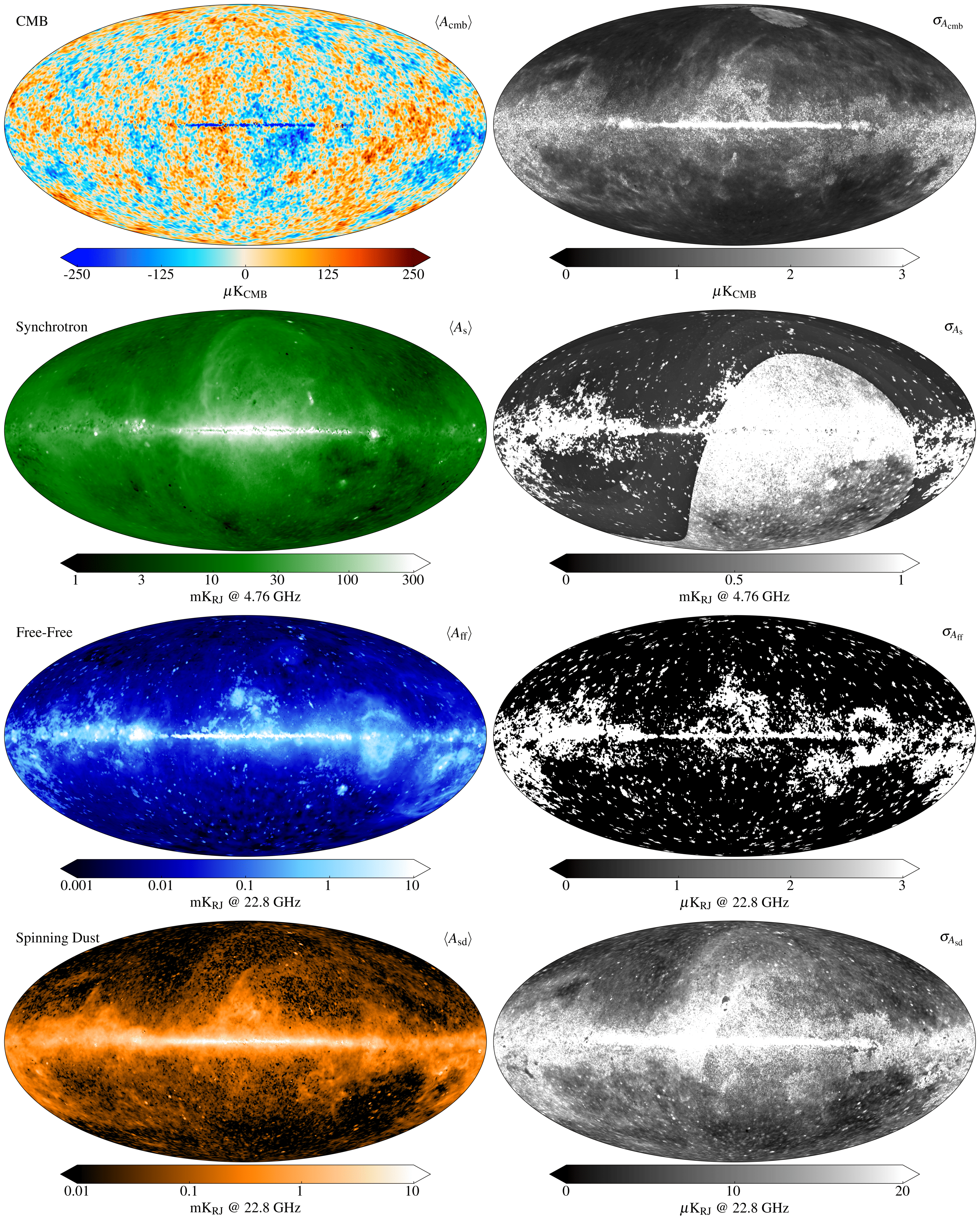}
    \caption{Maximum likelihood maps (left) and standard deviations (right) for the CMB, synchrotron, free--free, and spinning dust components at their reference frequencies, shown in Galactic Mollweide projection. Foreground means are displayed in logarithmic scale, while the CMB and all standard-deviation maps use a linear scale.}
    \label{fig:lf_foregrounds}
\end{figure*}

\subsection{CMB AND GALACTIC FOREGROUNDS}\label{subsec:cmbfore}
We now present the results for the CMB and the three low-frequency foreground components constrained by our analysis. Figure~\ref{fig:lf_foregrounds} shows the maximum likelihood amplitude maps together with their associated posterior standard deviations. To highlight the impact of the revised low-frequency modeling, Fig.~\ref{fig:diff_lf_foregrounds} compares our maximum likelihood amplitudes with those obtained from the \textit{Planck} 2015 model at a resolution of $\mathrm{FWHM}=2^\circ$ to highlight the large-scales. The corresponding spectral parameters are shown in Fig.~\ref{fig:lf_foregrounds_parameters}.

\subsubsection{Cosmic Microwave Background}\label{sec:res_cmb}
The CMB signal is constrained by a joint fit to WMAP and \textit{Planck} data, with the \textit{Planck} HFI channels providing the dominant statistical weight. As the CMB modeling and data combination are unchanged relative to the \textit{Planck} 2015 model, no major differences in the recovered CMB component are expected. The top-left panel of Fig.~\ref{fig:lf_foregrounds} shows the posterior mean CMB map derived in this work. The main deviations relative to the \textit{Planck} 2015 solution are confined to the Galactic plane, as shown in the top-left panel of Fig.~\ref{fig:diff_lf_foregrounds}. While differences at high Galactic latitudes remain minimal, deviations reach up to $\sim25\%$ in the inner Galactic plane and $\sim5\%$ in regions such as the Gum Nebula and Eridanus. These differences originate from a revised partitioning of emission among the low-frequency foregrounds, in particular synchrotron and free--free emission below $\sim20$\,GHz. Owing to the relatively flat free--free spectrum ($\beta_{\mathrm{ff}} \approx -2.1$), its contribution remains non-negligible up to frequencies approaching $\sim100$\,GHz in the Galactic plane, and residual component degeneracies can indirectly affect the recovered CMB signal in these regions.

Consistent with this interpretation, the top-right panel of Fig.~\ref{fig:lf_foregrounds} shows an increase in the CMB posterior uncertainty in the Galactic plane relative to high latitudes. Despite these localized differences, the large-scale statistical properties of the CMB remain unchanged. Using the LM95 mask (see Section~\ref{sec:gof}), the $C_\ell$ spectra agree to within 2\% over the multipole range $\ell = 2$--$180$, indicating no significant large-scale deviation from the \textit{Planck} 2015 model. For reference, the maximum-likelihood amplitude of the kinetic quadrupole template fitted in the LFI 70\,GHz channel is 0.53 relative to its nominal normalization.

\subsubsection{Synchrotron}\label{sec:res_sync}
The synchrotron component is the foreground most affected by the modeling adopted in this work. Figure~\ref{fig:lf_foregrounds} shows the posterior mean synchrotron amplitude at 4.76\,GHz, where the main large-scale Galactic structures and spurs, including the North Polar Spur are clearly recovered. The spatial morphology is consistent with previous analyses, indicating that the dominant difference relative to the \textit{Planck} 2015 model lies in the overall normalization rather than in the morphology of the emission. This is quantified through a T--T analysis between our synchrotron map and the \textit{Planck} 2015 synchrotron prediction evaluated at 4.76\,GHz, performed at $\mathrm{N_{\textrm{SIDE}}}=64$, masking unreliable regions with the LM95 mask and excluding compact sources using the LFI\,30\,GHz point-source mask (PS30; \citealt{planck2014-a03}). The resulting Spearman rank coefficient of $r_{\mathrm{s}}=0.94$ confirms the strong morphological agreement, while the best-fitting slope of $0.47\pm0.02$ indicates that the \textit{Planck} 2015 model underestimates the synchrotron amplitude at 4.76\,GHz by approximately 50\%. This offset arises from the revised low-frequency modeling adopted here, which anchors the synchrotron component directly at 2.30 and 4.76\,GHz instead of extrapolating from 408\,MHz. A T--T analysis between C-BASS and the Haslam 408\,MHz map yields a spectral index of $\beta_{0.408-4.76}=-2.82\pm0.02$, flatter than the $\beta=-3.1$ assumed over the same frequency range in the \textit{Planck} 2015 model. The resulting spectral difference of $\Delta\beta_\mathrm{s}=-0.26\pm0.02$ implies an amplitude shift of approximately 50\% between 0.408 and 4.76\,GHz, fully accounting for the normalization offset observed in the T--T comparison. Since this comparison is restricted to high Galactic latitudes, where synchrotron emission is expected to dominate, the flatter effective slope is unlikely to be driven by free--free contamination. Moreover, the combined absolute calibration uncertainty of Haslam and C-BASS is at the level of $11\%$, assuming 10\% for Haslam and 5\% for C-BASS added in quadrature, and cannot account for the $\sim 50\%$ normalization difference implied by adopting $\beta_\mathrm{s}= -3.1$.

\begin{figure*}
	\includegraphics[width=\textwidth]{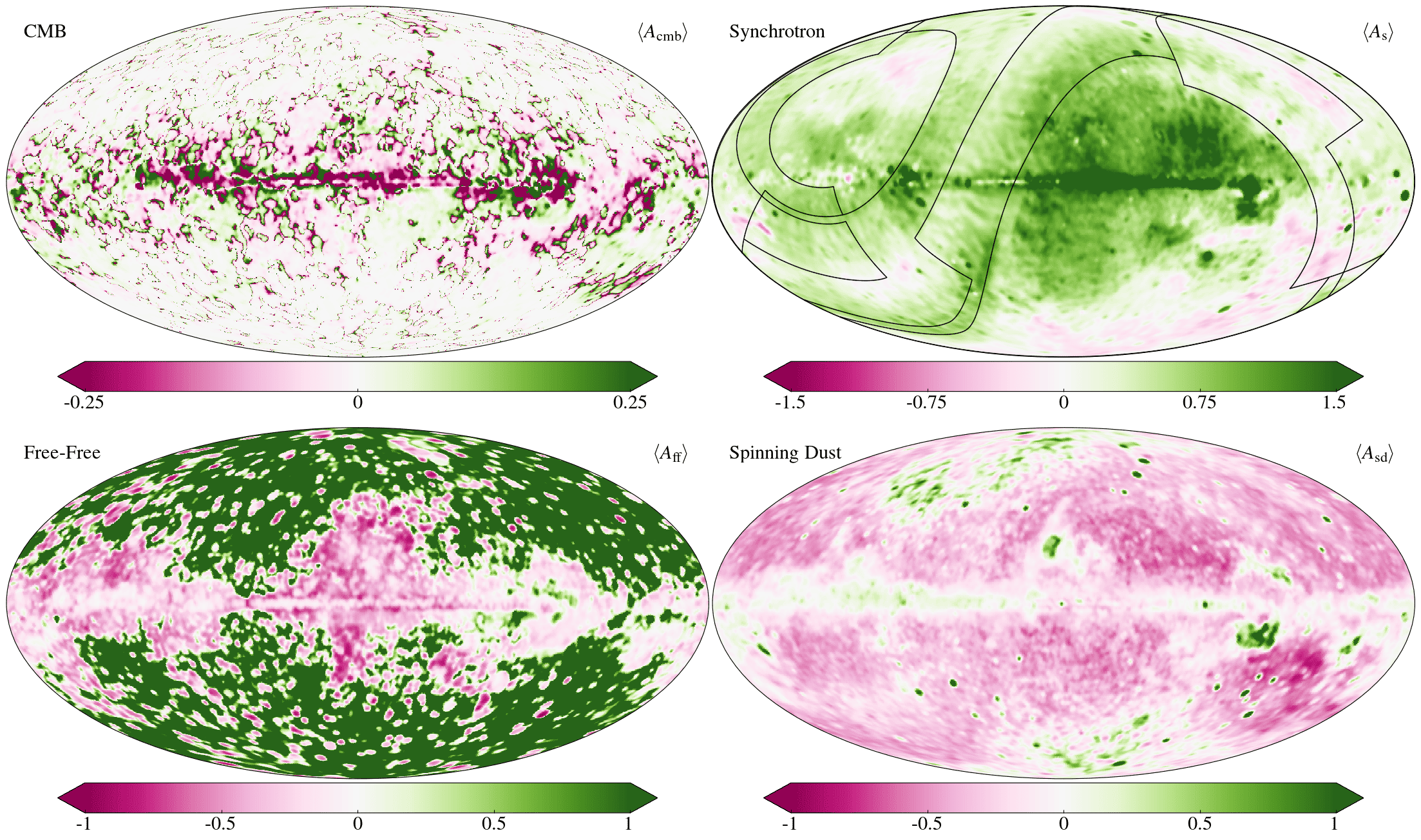}
    \caption{Fractional difference maps at $2^\circ$ resolution for the CMB and the low-frequency foregrounds highlighting large-scale modes. The fractional difference is defined as $\left(\langle A_{\text{This work}}\rangle - \langle A_{\text{\textit{Planck} 2015}}\rangle\right) / \langle A_{\text{\textit{Planck} 2015}}\rangle$. \textit{Top-left map:} CMB in thermodynamic units. \textit{Top-right:} Synchrotron at 4.76\,GHz with the Haslam atlas sky zones for the different surveys overlaid for reference \citep{Haslam1982}. \textit{Bottom-left:} Free--free at 22.8\,GHz. \textit{Bottom-right:} Spinning dust at 22.8\,GHz. The projection is Galactic Mollweide.} 
    \label{fig:diff_lf_foregrounds}
\end{figure*}

Beyond this modeling difference, the fractional-difference maps shown in Fig.~\ref{fig:diff_lf_foregrounds} provide evidence that the earlier solution was affected by large-scale systematics. Two independent signatures support this interpretation: (i) residual structures aligned with the Haslam scan strategy, which are identifiable by eye in the original unreprocessed 408\,MHz map, and (ii) sharp morphological boundaries that spatially coincide with the transitions between the individual surveys composing the Haslam atlas. These features are absent in our solution, indicating that the revised synchrotron map is largely free of such artifacts. The higher synchrotron normalization recovered here reflects both the improved low-frequency modeling and the mitigation of Haslam-related systematics. In the 20--40\,GHz range, the corresponding amplitude differences are typically\,3--15\,\si{\micro\kelvin}.

\begin{figure*}
	\includegraphics[width=\textwidth]{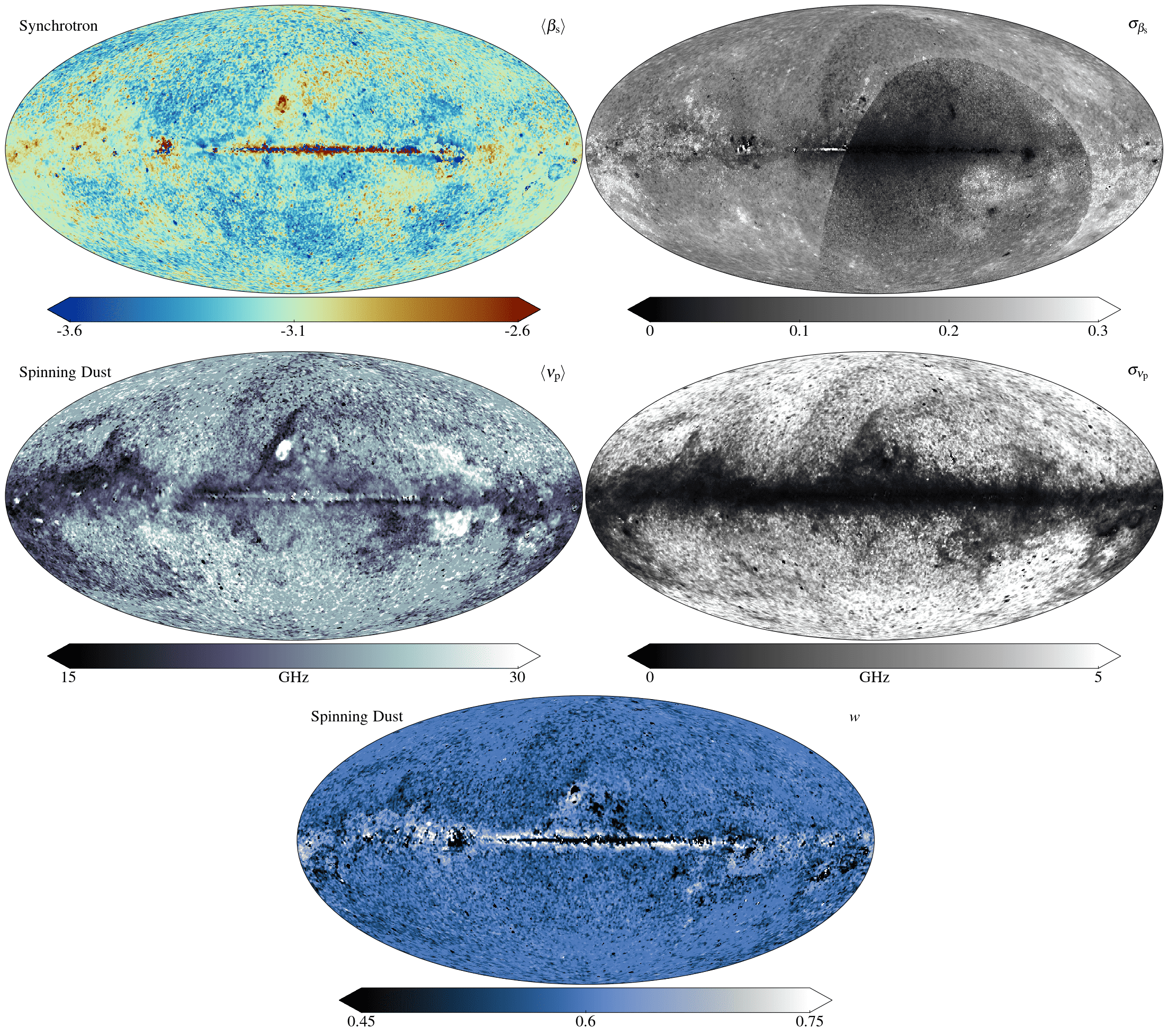}
    \caption{Maximum likelihood maps (left column) and corresponding posterior standard deviations (right column) for the synchrotron spectral index $\beta_\mathrm{s}$, the spinning dust peak frequency $\nu_\mathrm{p}$, and the spinning dust width $w$ maximum likelihood. The projection is Galactic Mollweide and all the scales are linear.}
    \label{fig:lf_foregrounds_parameters}
\end{figure*}

The synchrotron spectral index map provides an additional and independent validation of the revised low-frequency modeling. The map, shown in the top-left panel of Fig.~\ref{fig:lf_foregrounds_parameters}, is smooth across most of the sky and dominated by coherent morphological structures and has no scan-aligned or survey-dependent systematics. The global median spectral index is $\tilde{\beta}_{\mathrm{s}}=-3.18$, while a median weighted by synchrotron amplitude yields $\tilde{\beta}_{\mathrm{s}}=-3.17$, indicating that the brightest regions dominate the effective synchrotron spectrum inferred by template-fitting approaches. The overall sky dispersion of the spectral index is $\sigma_{\mathrm{sky}}\simeq0.13$. For the 4.76--22.8\,GHz baseline, a 5\% relative calibration uncertainty from C-BASS alone induces a spectral-index uncertainty of $\sigma_{\beta,\mathrm{cal}}\sim0.045$, while a $\sim1$\,mK monopole uncertainty at 4.76\,GHz contributes $\sigma_{\beta,\mathrm{mono}}\sim0.03$ in bright Galactic-plane regions and up to $\sim0.06$ at high Galactic latitudes where the synchrotron brightness temperature is lower. The combined systematic uncertainty is of order $\sim0.05$ in the plane and $\sim0.07$ at high latitudes, remaining subdominant relative to the observed spatial dispersion. The median posterior standard deviation of the spectral index within the LM95 mask is $\tilde{\sigma}_\beta\simeq0.16$, indicating that local uncertainties driven by component degeneracies and limited S/N exceed the calibration-induced systematic floor. A comparison between our synchrotron spectral index map and independent results from the literature is presented in Section~\ref{sec:si_comp}.

In compact regions where free--free emission dominates particularly around bright point sources at low-latitudes, strong degeneracies between components can drive the synchrotron amplitude toward zero, producing small artificial holes in the amplitude map. These pixels are corrected in a post-processing step by replacing near-zero values with the median of valid neighboring pixels within a $2^\circ$ radius. This procedure affects only a small fraction of the sky and has no impact on the large-scale synchrotron morphology or on the statistical properties of the spectral index map.

\subsubsection{Free--free}\label{sec:res_ff}
The free--free component is intrinsically faint over most of the sky, which makes its separation from synchrotron and spinning dust emission challenging outside lower latitudes. Figure~\ref{fig:lf_foregrounds} shows the posterior mean free--free amplitude and its associated uncertainty. At high Galactic latitudes, where free--free emission is weak, the solution is dominated by the dust-corrected H$\alpha$ template. In this regime, prominent structures such as the Gum Nebula and the Eridanus region are robustly recovered, including in the southern sky, and no significant leakage into the other low-frequency components is observed. This indicates that the separation of free--free emission is stable in the H$\alpha$-dominated regime at $1^\circ$ resolution.

The structure of the uncertainty map reflects the hybrid nature of the modeling. Although the FF47 mask formally allows pixel-by-pixel fitting over a larger area, the additional constraint that the fitted free--free brightness must exceed a fraction of the H$\alpha$ prediction restricts pixel-level fitting to the brightest regions. As a result $78\%$ of the sky is described by the H$\alpha$ template with a single fitted amplitude, while the remaining $22\%$ is determined through pixel-by-pixel fitting. Consequently, visible uncertainties are confined to the brightest free--free regions where pixel-by-pixel fitting is performed. In the remaining sky, where the free–free emission is described by the H$\alpha$ template with fixed morphology and amplitude, no posterior uncertainty is sampled for this component. A small number of compact features visible in the free–free amplitude map are likely associated with flat-spectrum radio sources, such as AGNs. Because the present sky model does not include a dedicated spectral component for flat-index sources, their emission is partially absorbed by the free–free term, whose spectral index ($\beta_\mathrm{ff}\approx-2.1$) is significantly flatter than that of diffuse synchrotron emission. This effect is confined to isolated compact regions and does not impact the large-scale diffuse free--free morphology.

The fractional difference between our free--free solution and the \textit{Planck}~2015 model is shown in the bottom-left panel of Fig.~\ref{fig:diff_lf_foregrounds}. At high Galactic latitudes, the fractional differences are off-scale because the \textit{Planck}~2015 free--free model tends toward zero in these regions, whereas the H$\alpha$-anchored solution presented here yields well-defined, non-zero values. In contrast, within the Galactic plane the free--free amplitude is systematically lower than in the \textit{Planck}~2015 model. This reduction is a direct consequence of the revised low-frequency modeling: increased flexibility in the synchrotron and spinning dust spectral descriptions reduces the need for free--free emission to absorb spectral mismatches, such that only emission consistent with the characteristic free--free index near $-2.1$ is retained in that component. In the inner plane ($|b|<1^\circ$), the free--free amplitude decreases by approximately $10\%$ relative to the \textit{Planck}~2015 model, a trend that is consistent with independent constraints from radio recombination line measurements \citep{planck2014-a31}. Overall, the revised solution yields a cleaner and more physically motivated free--free component in regions of strong component degeneracy. The nature of these degeneracies, however, differs between hemispheres owing to the distinct low-frequency anchors employed in each region. In the southern sky, S-PASS at 2.30\,GHz probes a regime where diffuse synchrotron emission strongly dominates over free–free, providing a comparatively clean synchrotron anchor. In the northern sky, C-BASS at 4.76\,GHz operates closer to the frequency range where free–free becomes non-negligible in the Galactic plane, increasing the intrinsic synchrotron–free--free degeneracy at that frequency. This effect is mitigated by the inclusion of QUĲOTE data at 11–19\,GHz.

\subsubsection{Spinning Dust}\label{sec:res_spdust}
The final low-frequency foreground component considered in this work is spinning dust emission. Figure~\ref{fig:lf_foregrounds} shows the posterior mean amplitude at 22.8\,GHz together with its uncertainty map. The morphology is dominated by diffuse Galactic structures and closely follows the large-scale dust distribution, indicating that the spatial structure is robustly recovered. A T--T comparison with the \textit{Planck}~2015 model at $\mathrm{N_{\textrm{SIDE}}}=64$, using the LM95 and PS30 masks, yields a Spearman coefficient of $r_{\mathrm{s}}=0.95$, confirming strong morphological agreement. Our solution is $\sim5\%$ fainter on average over the full sky and $\sim25\%$ fainter at high Galactic latitudes (PM61 mask), reflecting a modest renormalization driven by the revised low-frequency anchoring rather than structural differences. Figure~\ref{fig:lf_foregrounds_parameters} presents the posterior mean peak frequency $\langle\nu_{\mathrm{p}}\rangle$ and spectral width $w$. In the inner Galactic plane, peak frequencies lie predominantly in the range $25$--$30$\,GHz, falling within the $15$--$30$\,GHz range reported for individual sources by \cite{Fernandez2023} and \cite{Cepeda-Arroita2025}, while surrounding diffuse regions exhibit lower values of $15$--$20$\,GHz. In regions with $\mathrm{S/N}>10$, the typical value is $22\pm4$\,GHz with a median posterior uncertainty of\,$1$\,GHz. The width map is noisier, reflecting limited constraining power, but reproduces broader spectra in the Galactic plane on large angular scales. In the brightest regions, $w$ approaches the lower prior bound, signaling residual degeneracies with other low-frequency components. Allowing $w$ to vary spatially is central to the present modeling: unlike the \textit{Planck}~2015 model, which required two spinning dust components with fixed template shapes, the flexible width parameter captures the observed spectral diversity within a single phenomenological component.

Several structures illustrate the impact of the revised modeling. A diffuse loop closely following the outside of the North Polar Spur, mostly absent in the \textit{Planck}~2015 model, is clearly detected and spatially correlated with a thermal dust loop in the same region and was first noted by \cite{CepedaArroita2022PhD}. Its recovery is primarily enabled by C-BASS, whose frequency proximity to the spinning dust peak improves spectral leverage against synchrotron contamination (see \citealt{Dickinson2019}). The absence of this structure in the earlier model is likely attributable to limited low-frequency anchoring rather than weak intrinsic emission. The region where the Gum Nebula overlaps with the Vela supernova remnant, $(l,b)=(260^\circ,-3^\circ)$, does not exhibit enhanced spinning dust amplitude, but instead shows elevated peak frequencies. In the southern sky, S-PASS constrains the synchrotron amplitude at 2.30\,GHz, while the spinning dust spectrum is constrained mainly by WMAP and Planck LFI at higher frequencies. The large frequency gap between these data sets leaves more freedom for synchrotron--spinning-dust degeneracies than in the northern sky, where C-BASS provides an additional low-frequency anchor at 4.76\,GHz. It remains unclear whether the higher \(\nu_{\mathrm{p}}\) inferred in the Gum/Vela region is physical or reflects residual parameter degeneracies.

H\,\textsc{ii} $\zeta$ Ophiuchi, $(l,b)\approx(6^\circ,+24^\circ)$, is an $\approx 8^\circ$ diameter region powered by the nearby runaway O-type star $\zeta$\,Oph, comprising a diffuse ionized bubble together with associated dark and molecular clouds. The region is well known for harbouring a prominent bow shock $\approx1^\circ$ in extent, located only $\approx5'$ from $\zeta$\,Oph itself and detected clearly in both optical and infrared observations (e.g.~\citealt{Green2022}); however, this bow shock does not account for the bulk of the thermal dust or AME emission. Instead, most of the thermal dust and spinning-dust signal appears to originate from an extended dust filament ($\sim4^\circ$ long) at lower Galactic latitude, some $3$--$4^\circ$ away. The region exhibits elevated AME peak frequencies, with $\nu_\mathrm{p}=33\pm7$\,GHz on average and its north-western side approaching $\approx45$\,GHz. The complex geometry and strong free-free emission (accounting for $\approx95$--$98\%$ of the total emission at 30\,GHz) make component separation challenging and lead to elevated residuals (see Fig.~\ref{fig:residual_maps} in Section~\ref{sec:gof}), and may introduce component degeneracies, consistent with a similar spectral flattening seen in the synchrotron index across the same region. The anomalously high peak frequencies in the north-western corner may additionally be influenced by the base of the North Polar Spur, which runs adjacent to and possibly overlaps this part of the nebula. Nonetheless, the elevated peak frequencies are likely physically real: as discussed below, the region closely resembles the California and Tadpole nebulae in its physical environment, both of which independently exhibit high $\nu_\mathrm{p}$. The source was not included in \citet{Cepeda-Arroita2025} due to its large angular extent, but subsequent targeted aperture photometry has independently confirmed similarly high AME peak frequencies in this region (R.~Cepeda-Arroita, private communication).

Another source with an exceptionally high peak frequency is the California Nebula, $(l,b)\approx(160^\circ,-12^\circ)$, which reaches $\nu_\mathrm{p}=50\pm2$\,GHz in our map despite the Gaussian prior $N(25\pm5\,\mathrm{GHz})$ strongly disfavoring values above 40\,GHz. Without informative priors, \citet{Cepeda-Arroita2025} recovered $62\pm13$\,GHz for this region, indicating that the prior modestly suppresses the recovered peak frequency for extreme sources. In a Gaussian approximation, the prior itself would shift the inferred peak frequency downward by~$\sim5$\,GHz and would only slightly reduce the formal uncertainty on $\nu_\mathrm{p}$. This suggests that the substantially smaller error bars in our analysis arise mainly from the additional constraints provided by the AME-width prior and from explicitly fitting for the relative calibration between datasets, which is particularly important in regions where AME contributes only a small fraction of the total emission. Three further sources with $\nu_\mathrm{p}>40$\,GHz are identified in \citet{Cepeda-Arroita2025}: the Tadpole Nebula (IC\,410), Sharpless\,280, and W40. Of these, the Tadpole Nebula, $(l,b)\approx(173^\circ,-2^\circ)$, also appears as a high-$\nu_\mathrm{p}$ source in our map ($\approx40$\,GHz), and likely belongs to a similar physical environment as $\zeta$\,Oph and the California Nebula. Taken together, these results indicate that the peak-frequency map presented here modestly underestimates the true spectral complexity of AME for a minority of physically extreme sources, whose inferred $\nu_\mathrm{p}$ is partially lowered by the Gaussian prior. We do not expect the prior to bias the low-frequency end of the distribution in regions where AME is detected with significance: any AME component peaking below $\sim10$\,GHz would be increasingly faint and would likely fall below the sensitivity of the current data, as discussed in \citet{Cepeda-Arroita2025}. However, in intrinsically low-emission regions, particularly at high Galactic latitude, the recovered peak frequencies are increasingly prior-dominated and therefore may not accurately reflect the true underlying $\nu_\mathrm{p}$ distribution.

One important aspect of the $\nu_{\mathrm{p}}$ map is that, in several regions, the map shows a correspondence with dust morphology but an anti-correlation with the spinning dust amplitude. This is consistent with residual parameter degeneracies in the $10$--$40$\,GHz range, where the partition of signal between synchrotron, free--free, and spinning dust is sensitive to the adopted low-frequency anchoring. Part of this behavior may also be linked to the relative monopole choices adopted for the low-frequency anchor maps and the AME-sensitive WMAP/LFI channels. The anti-correlation is recovered consistently across internal tests with different priors and alternative monopole choices, indicating that it is not tied to a single specific configuration of the model.

\subsection{GOODNESS-OF-FIT}\label{sec:gof}
\begin{figure*}
    \centering
    \includegraphics[width=\linewidth]{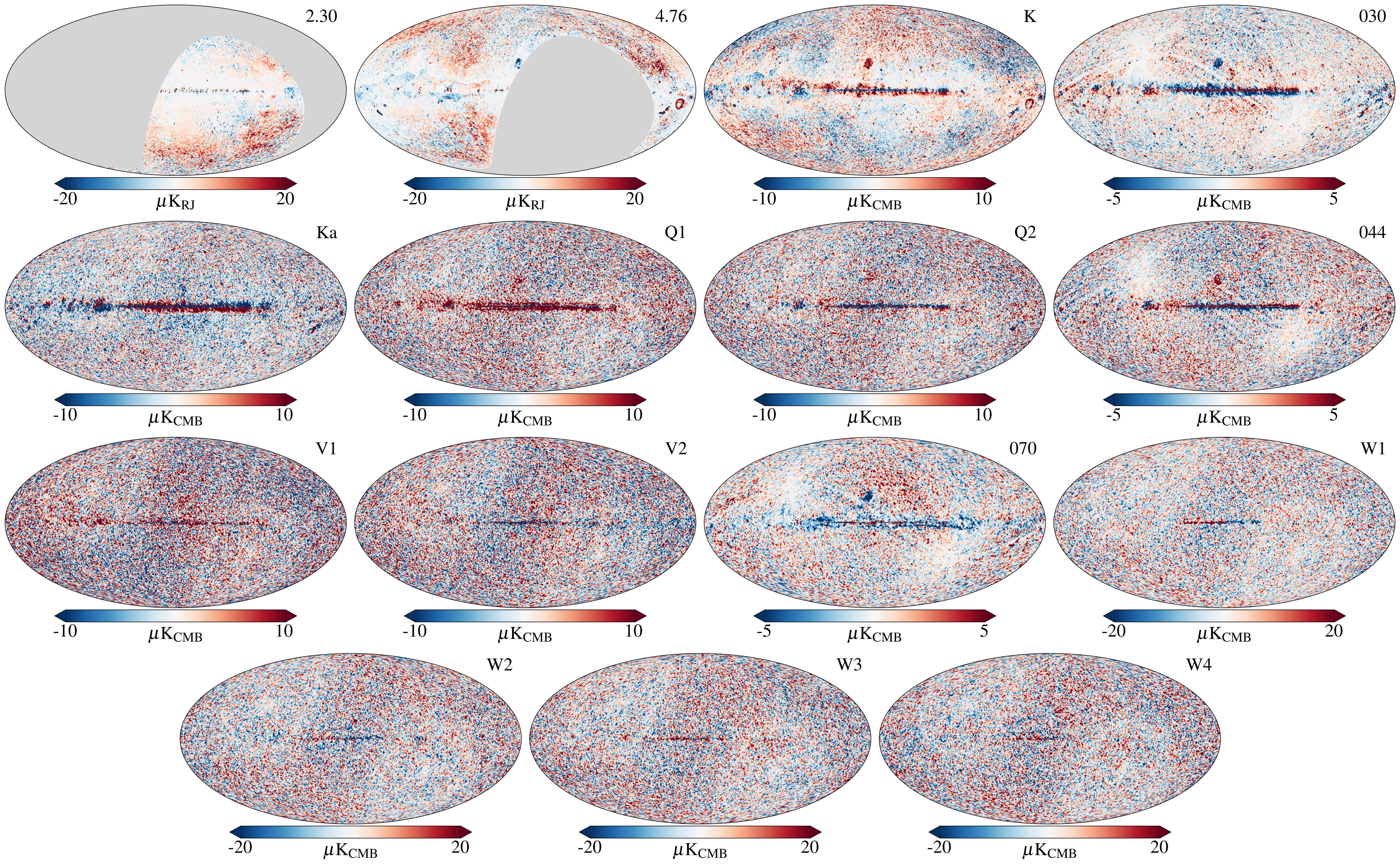}
    \caption{Residual maps, $d_{\nu}-s_{\nu}$, for each detector data set included in the baseline joint S-PASS, C-BASS, \textit{Planck} LFI and WMAP. All panels employ linear color scales. The label in the top-right corner of each panel indicates the frequency channel. All the employed scales are linear.}
    \label{fig:residual_maps}
\end{figure*}

\begin{figure}
	\includegraphics[width=0.48\textwidth]{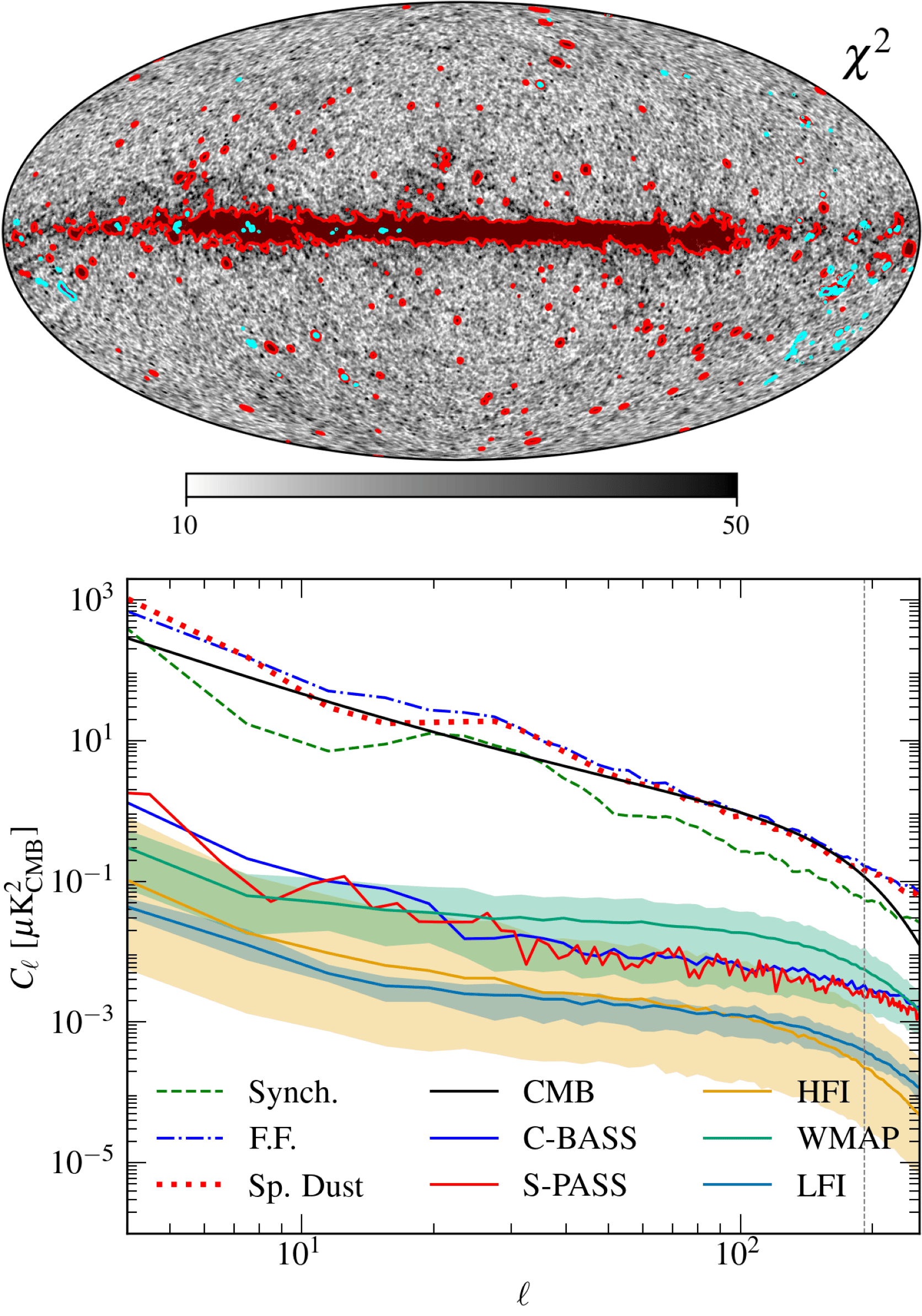}
    \caption{Goodness-of-fit diagnostics for the model. \textit{Top:} Map of $\chi^2$ per pixel at $1^\circ$ angular resolution. Red contours indicate the LM95 mask used in the analysis. Cyan contours indicate the regions where the Synchrotron map was inpainted in Fig~\ref{fig:diff_lf_foregrounds}. \textit{Bottom:} Angular power spectra of the residual maps for each frequency channel after subtraction of the best-fitting model. For each instrument class (LFI, HFI, and WMAP), the shaded regions show the minimum and maximum residual power at each multipole, while the solid lines indicate the corresponding mean spectra. The black curve shows the CMB angular power spectrum for reference. The dashed green, dash-dotted blue, and dotted red curves indicate the synchrotron, free--free, and spinning dust power spectra, respectively, evaluated at 30\,GHz.}
    \label{fig:gof}
\end{figure}

The goodness-of-fit analysis follows the same methodology adopted in the \textit{Planck}~2015 model (Sec. 5.3.1), allowing for a direct and consistent comparison. Residual maps are defined as
$\mathbf{r}_\nu = \mathbf{d}_\nu - \mathbf{s}_\nu$,
where $\mathbf{d}_\nu$ denotes the observed data map at frequency $\nu$ and $\mathbf{s}_\nu$ is the corresponding best-fitting sky model, including all astrophysical components and instrumental corrections. The residual maps are shown in Figs.~\ref{fig:residual_maps} and~\ref{fig:gof}, while quantitative goodness-of-fit metrics are summarized in Table~\ref{tab:gof}. As in the previous analysis, residuals are evaluated in brightness temperature units at a common angular resolution of $1^\circ$. The model provides an excellent fit over most of the sky, with residuals at high Galactic latitudes generally consistent with instrumental noise. Residuals in the \textit{Planck} HFI channels are not shown explicitly, as their behavior is essentially unchanged relative to the \textit{Planck} 2015 solution. The use of Cosmoglobe-reprocessed WMAP and LFI maps removes the large-scale CMB quadrupole residual present in the original \textit{Planck} 2015 LFI products, reducing typical LFI residual amplitudes from $\sim\pm10$\,\si{\micro\kelvin} to $\sim\pm5$\,\si{\micro\kelvin}. In the Galactic plane, residuals in the WMAP Q- and V-bands are reduced, while modest increases are observed in K-band and LFI\,30\,GHz in regions overlapping with QUIJOTE coverage. These changes reflect the revised balance of low-frequency foreground components when S-PASS and C-BASS are included. 

\input{tables/gof_stats}

The statistical goodness-of-fit is quantified through the $\chi^2$ map shown in the top panel of Fig.~\ref{fig:gof}, constructed from the Gaussian likelihood and summed over all frequency channels. As in the \textit{Planck} 2015 analysis, the effective number of degrees of freedom cannot be determined by simply counting parameters, owing to informative priors, spatial smoothing, and spatially constant instrumental parameters. Instead, we estimate the effective degrees of freedom by fitting a scaled $\chi^2$ distribution to the empirical $\chi^2$ distribution in clean regions of the sky. We adopt the conservative PM61 mask for this purpose and obtain an effective distribution corresponding to $\nu\simeq20$ degrees of freedom with a correlated-noise scaling factor of 1.09, identical to the value found in the \textit{Planck} 2015 model. Since 30 frequency channels enter the fit, this implies an effective number of $30-20=10$ free parameters per pixel, compared to 13 parameters in the formal model. The combined action of priors removes approximately three degrees of freedom, confirming that the solution remains predominantly data-driven. Based on this fit, we define the primary confidence mask LM95 by thresholding the $\chi^2$ map, smoothed with a $1^\circ$ Gaussian kernel, at $\chi^2=50$, corresponding to the upper tail of the distribution. This mask excludes approximately $5\%$ of the sky, and its complement defines the region where the likelihood is considered reliable. The LM95 mask is shown as the red contours in the top panel of Fig.~\ref{fig:gof}. Compared to the \textit{Planck}~2015 model, this represents an increase of $\sim3\%$ in usable sky fraction.

Additional diagnostics are provided by the residual angular power spectra and by channel-by-channel statistics. The bottom panel of Fig.~\ref{fig:gof} shows the residual $C_\ell$ spectra evaluated in the LM95 region, compared to the CMB and low-frequency foreground power spectra at 30\,GHz in the same sky fraction. Residual power remains at least two orders of magnitude below the CMB signal for multipoles $\ell<192$, demonstrating that the component separation is clean at the angular scales targeted in this work. Table~\ref{tab:gof} lists, for each channel, the RMS of the residual map outside the LM95 mask and its ratio to the corresponding instrumental noise RMS. For most channels this ratio is close to unity, indicating that residuals are consistent with noise not only in an integrated $\chi^2$ sense but also on a channel-by-channel basis. Compared to the \textit{Planck} 2015 model, S-PASS and C-BASS now dominate the synchrotron constraints at low frequencies, while the proximity of these surveys to WMAP K-band reduces the extent to which K-band alone dominated the low-frequency foreground solution in the earlier analysis. This leads to a modest redistribution of residual power across low-frequency channels without degrading the overall fit quality. For S-PASS and C-BASS, the $\sigma^{\mathrm{res}}_\nu/\sigma^{\mathrm{inst}}_\nu{}^{\mathrm{b}}$ values are $0.01$ and $0.03$, respectively. Such low ratios indicate that most of the stochastic variance of these channels is absorbed into the fitted large-scale model instead of remaining in the residual maps. For C-BASS, this sets a contamination floor of order $\sim170$\,\si{\micro\kelvin} in the recovered synchrotron amplitude map. Even so, the solution remains signal-dominated: in the faintest northern regions, after monopole subtraction, the synchrotron brightness is still of order $1$\,mK, corresponding to a contamination level of about $17\%$ in amplitude in the coldest parts of the sky. Some residual features are still visible in these two channels, including the weak anti-correlation seen in the Galactic spur, and are consistent with low-level local mismatches in the fitted spectral scaling of the low-frequency emission. Median fractional residuals within the LM95 region are below $0.2\%$ for all HFI channels, below $0.5\%$ for all LFI channels, below $1.5\%$ for all WMAP channels, and below $5\%$ for QUIJOTE channels, consistent with the adopted calibration uncertainties.

Overall, the goodness-of-fit diagnostics demonstrate that the revised model provides a statistically consistent description of the temperature sky. Residuals are noise-dominated over more than 95\% of the sky, the effective number of degrees of freedom matches that of the \textit{Planck} 2015 model, and the inclusion of more recent low-frequency surveys improves the physical consistency of the solution without introducing significant degradations in fit quality. These results confirm that the revised low-frequency modeling yields a robust component separation suitable for scientific interpretation.

\subsection{RECONSTRUCTED ALL-SKY MAP AT 4.76 GHz}
\begin{figure*}
	\includegraphics[width=\textwidth]{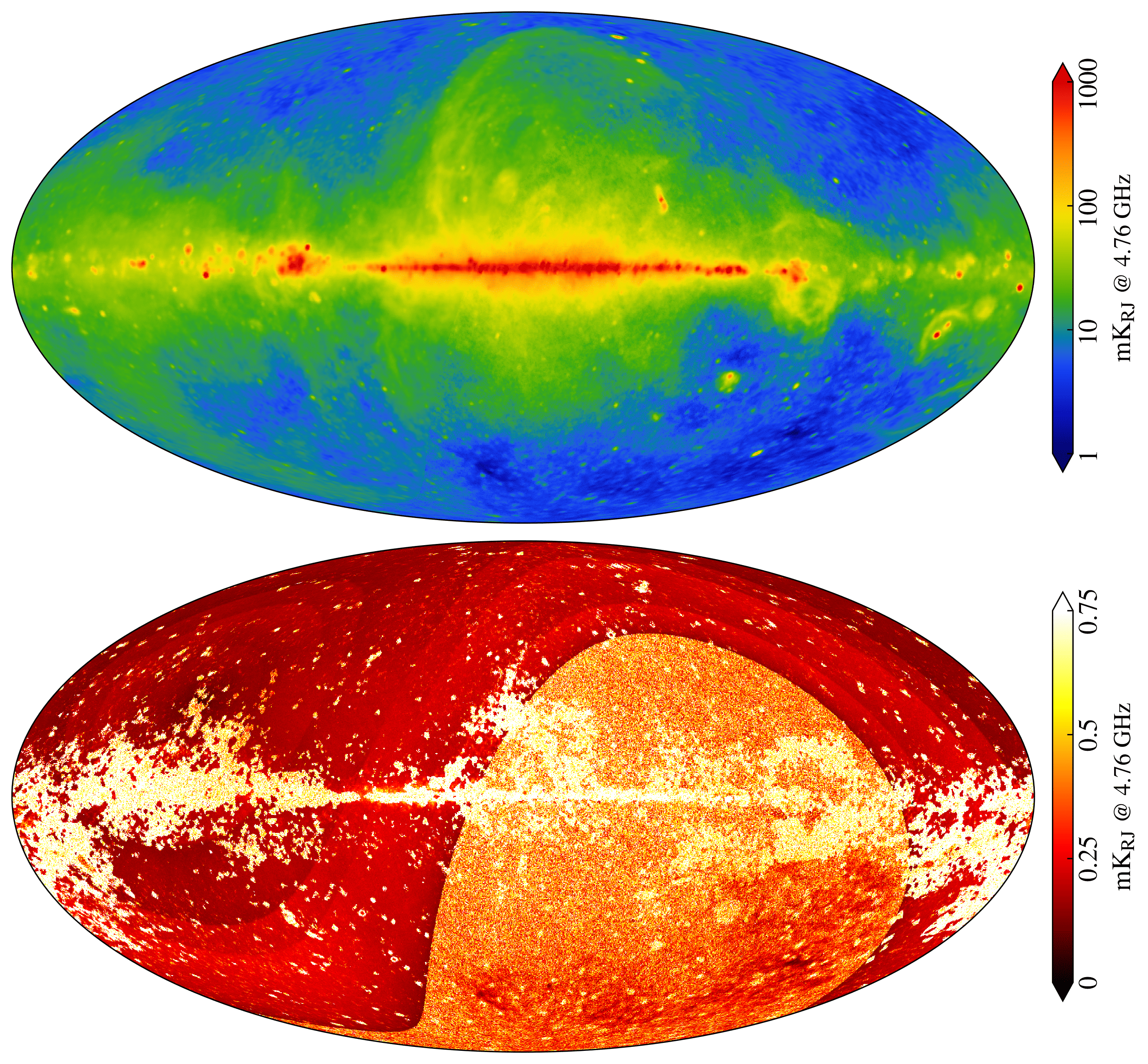}
    \caption{All-sky total-intensity map at 4.76\,GHz constructed from the maximum likelihoods presented in this work, shown together with its corresponding uncertainty map. The map is evaluated at $1^\circ$ angular resolution and represents the sum of the synchrotron, free--free, and spinning dust components at this frequency.}
    \label{fig:fs_5GHz}
\end{figure*}

Using the parametric model derived in this work, we generate a reconstructed all-sky total-intensity map at 4.76\,GHz by evaluating the fitted foreground components at this frequency. It should be interpreted as the best-fitting total-intensity sky model at 4.76\,GHz, combining the direct C-BASS constraint in the north with the model-based extrapolation of the southern low-frequency data. While the model can in principle be evaluated at any frequency, 4.76\,GHz represents the lowest frequency at which the solution is directly constrained over the full sky. Below this frequency, no other northern-sky data were used to generate this model, and any extrapolation increasingly relies on the assumed spectral model rather than on direct observational constraints. The resulting all-sky map, together with its associated uncertainty estimate, is shown in Fig.~\ref{fig:fs_5GHz}. The uncertainty map is obtained by propagating the posterior uncertainties of the individual components and summing them in quadrature after evaluating each component at 4.76\,GHz. At this frequency, the total intensity is dominated by synchrotron emission. Free--free emission is subdominant at 4.76\,GHz in the high-latitude regions considered here. In the C-BASS template-fitting analysis, \citet{Harper2022} estimate that free--free contributes less than \(20\%\) of the total 4.76\,GHz emission at high Galactic latitudes, while spinning dust and thermal dust are negligible. It is important to note that the northern sky at 4.76\,GHz is directly constrained by C-BASS observations, whereas the southern sky is obtained by extrapolating the S-PASS 2.30\,GHz map to 4.76\,GHz using the pixel-dependent spectral indices derived from the global parametric fit. As a consequence, the posterior standard deviation is systematically larger in the southern hemisphere, as visible in Fig.~\ref{fig:fs_5GHz}, reflecting the additional spectral extrapolation uncertainty. The boundary associated with the transition between surveys remains faintly visible by eye in the reconstructed map. These features reflect the non-uniform observational constraints entering the parametric fit in the transition from the C-BASS-constrained northern sky to the S-PASS-extrapolated southern solution.

\begin{figure}
    \centering
    \includegraphics[width=\linewidth]{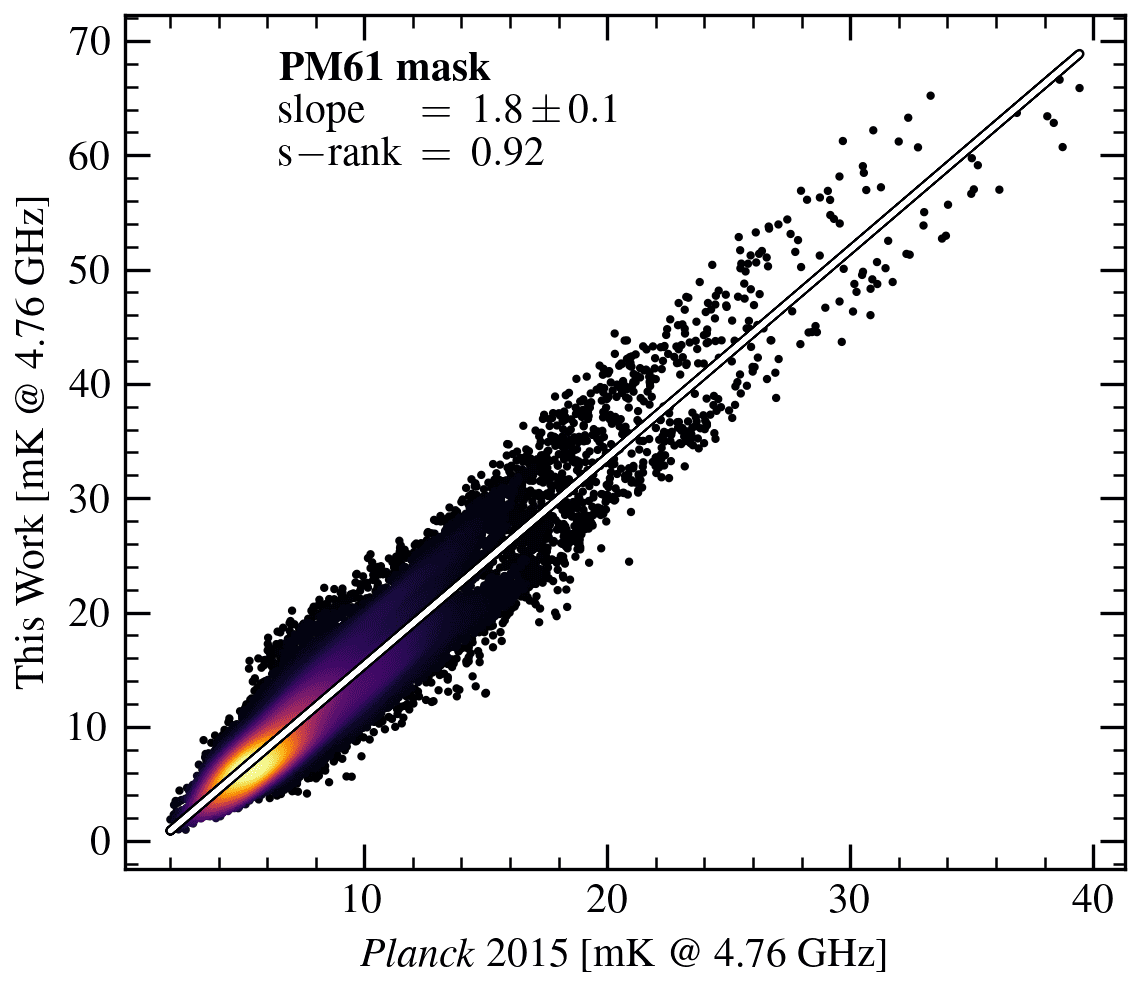}
    \caption{T--T plot between the \textit{Planck} 2015 model and the model presented in this work at 4.76\,GHz, evaluated at $\mathrm{N_{\textrm{SIDE}}}=64$ using the PM61 mask and the PS30 mask. The line shows the best-fitting linear relation.}
    \label{fig:psm_tt}
\end{figure}

A direct comparison with the \textit{Planck}~2015 model evaluated at the same frequency is presented through a T--T analysis shown in Fig.~\ref{fig:psm_tt}. The comparison is performed at $\mathrm{N_{\textrm{SIDE}}}=64$, masking the Galactic plane with the PM61 mask and excluding compact sources using the PS30 mask. The two models are highly correlated morphologically, confirming that they trace the same large-scale Galactic structures. However, the best-fitting slope departs significantly from unity, indicating that the \textit{Planck}~2015 model underestimates the total sky brightness at 4.76\,GHz by $\sim55\%$ relative to the present model, driven primarily by differences in the synchrotron component. The T--T distribution shows a clear bifurcation around the best-fitting relation: the upper branch is dominated by Galactic-plane pixels, while the lower branch is associated with high-latitude structures, most prominently the North Polar Spur. This behavior reflects the impact of allowing spatial variations in the synchrotron spectral properties across the sky, rather than assuming a spatially uniform spectral index as in the \textit{Planck}~2015 model. In the northern sky, the resulting map closely follows the original C-BASS observations. The main enhancement occurs in the southern hemisphere: instead of adopting a spatially uniform rescaling of S-PASS with monopole and dipole corrections, the present approach propagates a pixel-dependent synchrotron spectral index derived from the global parametric fit. This allows spatial variations in spectral shape to be consistently mapped to 4.76\,GHz, yielding a more physically realistic representation of the large-scale emission.

All-sky maps constructed in this way are particularly valuable for sky-simulations such as PSM, PySM, and related models. The 4.76\,GHz map provides a physically motivated intermediate-frequency anchor between traditional low-frequency radio surveys and microwave data, substantially reducing the need for long extrapolations from 408\,MHz or for purely phenomenological prescriptions. As such, it offers a robust input for simulations of Galactic foregrounds in the radio and microwave frequencies. Although 4.76\,GHz is the lowest frequency at which the present solution is directly constrained over the entire sky, extrapolation of the model to slightly lower frequencies remains feasible. In the northern sky, no data below 4.76\,GHz are included in the present parametric fit, and predictions at lower frequencies rely on spectral extrapolation of the fitted components rather than on additional observational constraints. Given that the inferred synchrotron curvature is small down to frequencies of order 1\,GHz, approximate all-sky maps can be constructed at 2.30\,GHz within the adopted spectral model.

\section{DISCUSSION}\label{sec:discussion}
This section addresses the physical interpretation of the reconstructed foreground components. The recovered maps and spectral parameters are compared with independent literature results, and the main limitations related to data coverage, calibration, and modeling assumptions are discussed. A more detailed investigation of each foreground component and a comprehensive comparison with independent studies will be the subject of future work.

\subsection{LOW FREQUENCY SPECTRAL INDEX}\label{sec:si_comp}
We compare the synchrotron spectral-index map derived in this work with previous all-sky reconstructions from the literature, while emphasizing that these products are not direct counterparts of our gigahertz-frequency solution. Most previous maps use longer frequency baselines, typically extending to 408\,MHz or below, whereas our synchrotron index is constrained by data above 2\,GHz. In the sub-gigahertz regime, the synchrotron spectrum is expected to flatten \citep[e.g.][]{Strong2011}, and optically thick free--free emission becomes increasingly important at long wavelengths, especially toward the Galactic plane. Differences between our map and low-frequency spectral-index reconstructions therefore reflect both genuine frequency-dependent astrophysical effects and residual systematics in the input surveys.

Together with the maximum likelihood synchrotron spectral index $\beta_\mathrm{s}$, spectral indices for the total sky emission can also be predicted from our model by evaluating the sky solution at two frequencies. Appendix~\ref{app:si} presents examples of such spectral-index maps. For reference, we compare our synchrotron spectral-index solution with the \citet{Irfan2023} map between 0.408 and 23\,GHz and with the \citet{Irfan2026} map, obtained by extrapolating their 43\,MHz--4.76\,GHz spectral model using their curvature map. We also compare spectral-index maps predicted from our model over selected frequency baselines with the maps of \citet{Giardino2002} and \citet{Platania2003}. All comparisons are performed through T--T analyses after smoothing to a common \(10^\circ\) resolution and degrading to $\mathrm{N_{\textrm{SIDE}}}=16$. We perform the analysis with the PM61 and LM95 masks, masking compact sources with the PS30 mask. The resulting Spearman rank coefficients are listed in Table~\ref{tab:t-t}.

All the literature spectral-index maps show weak pixel-by-pixel correlations with our solution. This should not be interpreted purely as a physical disagreement between reconstructions. Spectral-index maps are logarithmic intensity ratios and are therefore highly sensitive to calibration uncertainties, zero-level offsets, residual striping, and frequency-dependent foreground contamination. Moreover, the different frequency baselines probe different effective synchrotron spectra. The largest coefficient is obtained for the \citet{Giardino2002} map within the PM61 mask, but even in this case the correlation is only $r_{\mathrm{s}}=0.27$. Thus, none of the maps in the literature show a significant morphological match to our gigahertz-frequency synchrotron-index solution. We interpret Table~\ref{tab:t-t} as a diagnostic of differences in frequency coverage, data selection, and modeling assumptions, rather than as a direct test of the physical validity of previous reconstructions.

\input{tables/si_t-t_plots}

An additional limitation of the reconstructed synchrotron spectral-index map is the presence of dust-correlated structure. Some of the features visible in Fig.~\ref{fig:lf_foregrounds_parameters} resemble nearby-dust morphology (e.g. \citealt{Edenhofer2024}), suggesting that part of the large-scale variation in $\beta_{\mathrm{s}}$ is driven by residual degeneracy between synchrotron and spinning dust emission. This degeneracy was first pointed out by \citet{bennett2003b} and is supported here by the pairwise spectral-index maps in Appendix~\ref{app:si}, where combinations involving the QUIJOTE frequency range appear more dust-like, while the 2.30--4.76\,GHz map appears least affected by dust-correlated residuals, apart from obvious free--free dominated regions. To quantify this effect, we computed the Spearman rank correlation coefficient within the LM95 region between $\beta_{\mathrm{s}}$ and the nearby-dust template of \citet{Edenhofer2024}. The maximum-likelihood $\beta_{\mathrm{s}}$ map yields $r_{\mathrm{s}} = 0.32$, whereas the posterior-mean map gives $r_{\mathrm{s}} = 0.62$. This indicates that the dust-correlated structure becomes more pronounced in the posterior-mean reconstruction, although the correlation remains only moderate even in that case.

\subsection{DUST CORRELATION}\label{sec:dust_correlation}
\begin{figure}
    \centering
    \includegraphics[width=\linewidth]{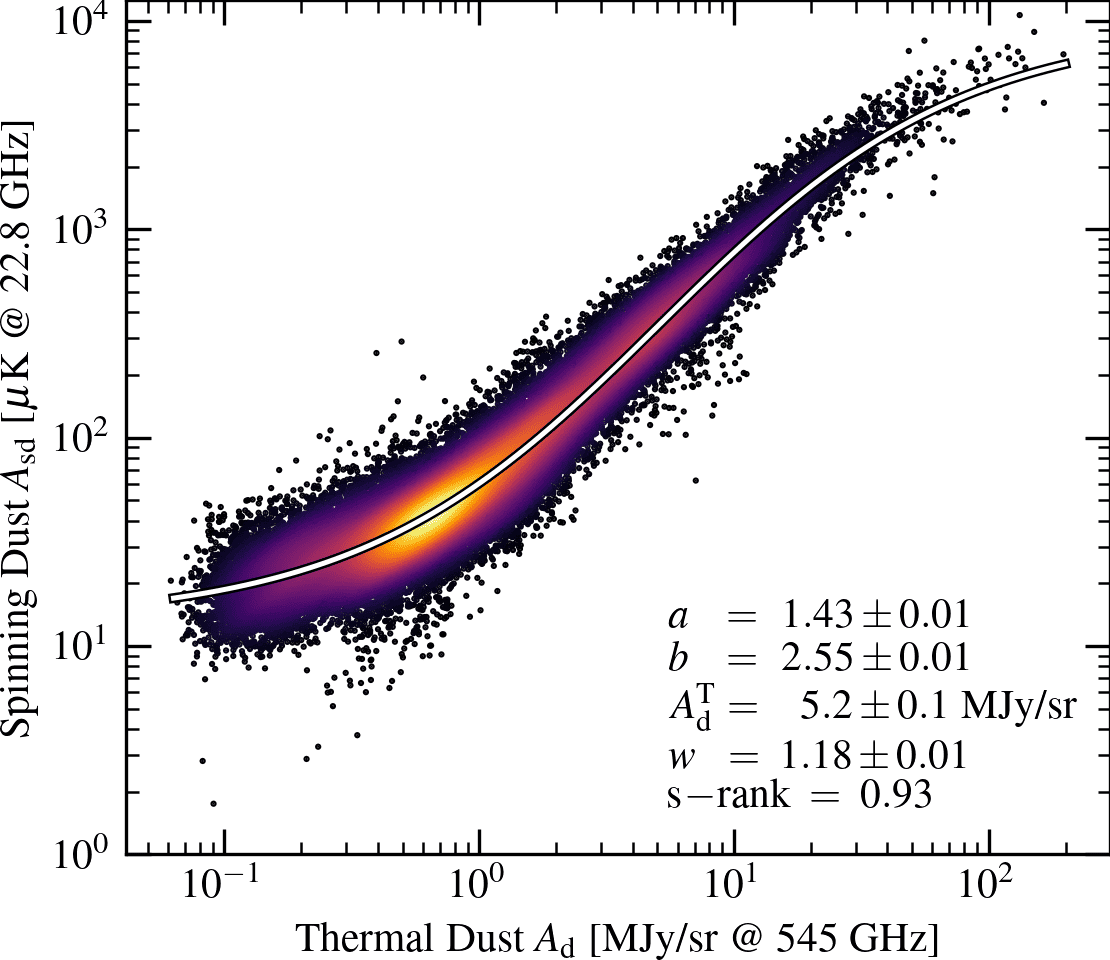}
    \caption{T--T plot between the spinning dust amplitude at 22.8\,GHz and the thermal dust amplitude at 545\,GHz, evaluated at a common angular resolution and $\mathrm{N_{\textrm{SIDE}}}$\,$=64$. Pixels are selected using the $\mathrm{S/N}>3$ mask \citep{planck2014-a31} and the PS30 mask. The color scale indicates the density of the scatter points, while the solid curve shows the best-fitting hyperbolic-tangent model.}
    \label{fig:dust_correlation}
\end{figure}

The spatial correlation between spinning dust and thermal dust emission reflects the link between electric-dipole emission from rapidly rotating very small grains and the overall dust column density. We quantify this relation through a T--T analysis between the spinning dust amplitude at 22.8\,GHz, $A_\mathrm{sd}$, and the thermal dust amplitude at 545\,GHz, $A_\mathrm{d}$, derived from our component-separation model. The distribution is shown in Fig.~\ref{fig:dust_correlation}. The analysis is performed at $\mathrm{N_{\textrm{SIDE}}}=64$, masking compact sources with the PS30 mask and applying the same $\mathrm{S/N}>3$ mask used in \cite{planck2014-a31}. A linear fit yields a slope of $74\pm7$~\si{\micro\kelvin}/MJy\,sr$^{-1}$ and a Spearman rank coefficient of $r_\mathrm{s}=0.93$. We also compare $A_\mathrm{sd}$ with the dust optical depth $\tau_{353}$ \citep{planck2013-p06b}, obtaining $r_\mathrm{s}=0.93$ and a best-fitting slope of $(12\pm1)\times10^{6}$\;\si{\micro\kelvin}/$\tau_{353}$. Since $\tau_{353}$ traces dust column density more directly than $A_\mathrm{d}$, being less sensitive to temperature variations, this provides a complementary test of whether spinning dust scales with dust mass rather than dust heating conditions. The slopes reported in \cite{planck2014-a31} are slightly lower than those obtained here because their correlation was computed using only the lower-frequency spinning dust component of the \textit{Planck} 2015 model. When the sum of both spinning dust components in the \textit{Planck} 2015 model is used instead, the inferred slopes increase and become slightly larger than ours by approximately $5\%$, consistent with the relative amplitude difference between the two solutions.

As shown in Fig.~\ref{fig:dust_correlation}, the T--T distribution departs from strict linearity at high thermal dust amplitudes, where the emissivity decreases. This behavior was already visible in \cite{planck2014-a31} and reflects reduced spinning dust efficiency in regions of high dust column density. To capture this transition, we model the T--T relation in logarithmic space using a hyperbolic tangent,
\begin{equation}
    \log A_{\rm sd} = a \, \tanh\!\left(\frac{\log A_{\rm d} - \log A_{\rm d}^{\rm T}}{w}\right) + b ,
\end{equation}
which describes two asymptotic regimes connected by a smooth transition and provides a more realistic representation of the full dynamic range than a single linear slope. The fitted transition parameter is \(A_{\rm d}^{\rm T} = 5.2\pm0.1\,\mathrm{MJy\,sr^{-1}}\), which marks the characteristic thermal dust amplitude where the spinning dust emissivity begins to flatten. The fit yields $\chi^2/\mathrm{dof}=133.7$, significantly larger than unity, as expected for a phenomenological model that does not attempt to capture the full intrinsic scatter of the data. Introducing an additional dispersion term to account for intrinsic variation requires a fractional scatter of $\sigma_\mathrm{scatter}=1.37$, corresponding to a $37\%$ dispersion. Physically, this transition scale indicates the dust column density above which environmental effects modify the small-grain population: at low to intermediate column densities, spinning dust efficiently traces the abundance of very small grains, while at higher column densities, grain growth, coagulation, and mantle formation reduce the relative population of the smallest rapidly rotating particles responsible for electric-dipole emission. Thermal dust emission continues to scale with total dust mass, leading to a decrease in the effective \(A_{\rm sd}/A_{\rm d}\) ratio beyond the transition. The hyperbolic-tangent model provides a compact phenomenological description of how spinning dust efficiency evolves with dust environment while preserving the strong overall dust correlation.

\subsection{ROLE OF 10--20~GHz DATA IN CONSTRAINING FOREGROUND SPECTRA}
\begin{figure}
    \centering
    \includegraphics[width=\linewidth]{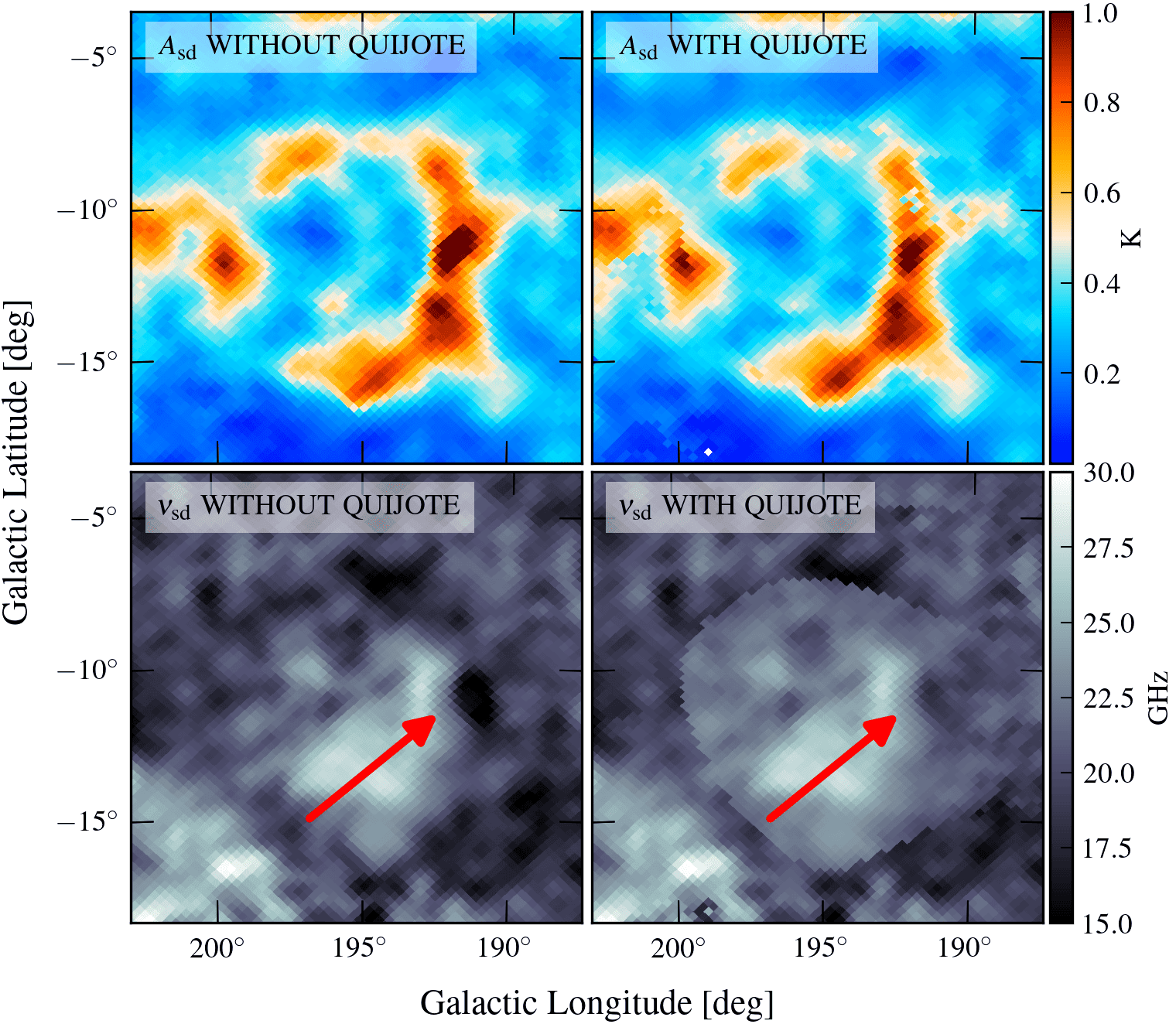}
    \caption{Comparison of spinning dust parameters in the $\lambda$~Orionis region. \textit{Top row}: spinning dust amplitude \(A_{\rm sd}\) at 22.8\,GHz without (left) and with (right) QUIJOTE data included in the fit. \textit{Bottom row}: corresponding peak frequency maps \(\nu_{\rm sd}\) without (left) and with (right) QUIJOTE. All panels show the same sky region in Galactic coordinates. The red arrow marks the position of the B30 cloud for reference.}
    \label{fig:lambda_ori}
\end{figure}

The 10--20\,GHz range is crucial for resolving degeneracies between synchrotron, free--free, and spinning dust emission, since it samples the transition from synchrotron-dominated emission to the rising side of the spinning dust spectrum, where the relative slopes of the components differ most strongly \citep{Poidevin2023}. Although this frequency range does not by itself fully bracket the spinning dust peak, it provides essential leverage on the rising part of the spectrum and, in combination with WMAP and \textit{Planck} channels above 20\,GHz, improves constraints on the peak frequency and spectral width. In our analysis, QUIJOTE data are used exclusively in the Galactic plane, where they contribute to constraining the spectral parameters of the emission components. Even with reduced statistical weight, the QUIJOTE channels provide anchoring in a frequency range that is otherwise sparsely covered, thereby reducing degeneracies in the component separation solution. Frequencies below this range mainly constrain synchrotron behavior, while those above 20--30\,GHz probe the declining side of the spinning dust spectrum together with free--free emission. A clear illustration is provided by the B30 cloud in the $\lambda$~Orionis region in Fig.~\ref{fig:lambda_ori}. When QUIJOTE data are excluded, the fit between C-BASS and WMAP/\textit{Planck} channels is weakly constrained on the rising side of the spinning dust spectrum, and the solution converges to a peak frequency near $\sim12$\,GHz. After including the 11 and 13\,GHz QUIJOTE channels, even with the 5\% calibration error added to the RMS uncertainties, the spectral solution shifts toward a peak near $\sim23$\,GHz, consistent with previous measurements of the region \citep{Harper2025,Cepeda-Arroita2025}. This example shows that coverage in the 10–-20,GHz range is essential for breaking degeneracies between synchrotron, free–free, and spinning dust, enabling more robust constraints on their spectral parameters and, potentially, on the width of the spinning dust spectrum. It underscores the need for future full-sky surveys in this frequency (e.g. \citealt{delaHoz2026} and \citealt{Hoyland2022}).

\subsection{MODEL LIMITATIONS}
Although the model presented here provides an improved description of the low-frequency Galactic sky relative to previous analyses, its interpretation remains limited by the available data, frequency coverage, and adopted parametric assumptions. A major advance of this work is the replacement of the 408\,MHz synchrotron anchor with S-PASS and C-BASS, reducing the need to connect widely separated frequency regimes with an explicitly curved synchrotron model. Even with these data, the separation between synchrotron spectral behavior, free--free emission, and the rising side of the spinning dust spectrum is not uniquely determined in all regions. The inferred spinning dust peak frequency and width, as well as the relative partition of emission between components, depend to some degree on the adopted spectral forms and priors, particularly in areas with incomplete 10–20\,GHz coverage or low S/N. These parameters should be regarded as effective descriptors of the observed spectrum rather than direct measurements of intrinsic grain properties. Fully breaking these residual degeneracies requires improved large-area surveys in the few-GHz to 20\,GHz range, providing robust coverage of both the synchrotron-dominated regime and the rising side of the spinning dust spectrum.

Additional limitations arise from survey characteristics, frequency coverage, and calibration uncertainties. The QUIJOTE maps, while crucial for constraining spectral behavior in the Galactic plane, are used here only over the brightest regions, since residual RFI and large-scale filtering limit the robustness of the total-intensity data at high Galactic latitudes. Outside the plane, the spinning dust spectrum is driven primarily by WMAP and \textit{Planck} LFI data together with model priors, and constraints in this range remain indirect. We have not attempted to model the sky below the frequencies of our anchor surveys, S-PASS and C-BASS. Extrapolations outside this range are therefore not directly data-driven and may be biased in regions with complex low-frequency spectral behavior. Furthermore, the derived synchrotron spectral index is sensitive to the adopted relative offsets between frequency channels. The uncertainties quoted in this work refer to the monopole and dipole offsets within the relative zero-level convention of the \texttt{Commander} analysis, not to the uncertainty in the absolute sky zero level. In particular, they do not include a possible absolutely calibrated, spatially smooth radio-background component that would have been removed by the map-making and offset conventions. Within this relative framework, monopole-offset uncertainties propagate to $\sigma_{\beta_\mathrm{s}}\sim0.05$ in the Galactic plane and up to $\sim0.07$ at high Galactic latitudes; these values should therefore be interpreted as uncertainties on the fitted relative synchrotron spectral index, rather than on an absolute sky spectral index.

The adopted component models also introduce intrinsic limitations. Synchrotron emission is described by a single power-law SED per pixel over the fitted range. Residual degeneracy between synchrotron and spinning dust can still lead to leakage between the two components, especially in regions where the overall spectrum is relatively flat or where the 10--20\,GHz constraints are weak. The inclusion of S-PASS and C-BASS provides important leverage on the synchrotron-dominated side of the spectrum, but does not completely remove this model dependence. This degeneracy is reflected morphologically in the recovered synchrotron spectral-index map, which shows moderate dust-correlated structure on large angular scales. The inferred $\beta_{\mathrm{s}}$ is an effective synchrotron-like spectral index within the adopted parametric model. Free--free emission outside the Galactic plane is largely template-driven through H$\alpha$, making the solution sensitive to dust-extinction corrections, H$\alpha$ scattering, and the adopted H$\alpha$--to--free--free conversion, while inside the plane the separation relies on fixed spectral assumptions and remains degenerate with the other low-frequency components. The spinning dust spectrum is represented by a single log-normal form, which is flexible enough for the available data but remains a phenomenological description. Real spinning dust likely arises from multiple environments, and the fitted peak frequency and width do not map uniquely onto specific grain populations.

Finally, the analysis is performed at a common angular resolution of $1^\circ$ FWHM, so compact H\,{\sc ii} regions and small-scale spinning dust sources are beam-diluted and their parameters represent large-scale averages rather than intrinsic source properties. Compact synchrotron-dominated sources with unusually flat spectra, such as Tau~A, are not well described by the diffuse synchrotron model and can produce localized residuals in amplitude and spectral index maps. Pixel-by-pixel fitting also does not fully correct for beam-shape differences between surveys. Residual beam asymmetries and scan-dependent systematics lead to small morphological mismatches of point sources between channels, producing characteristic quadrupolar residual patterns around bright sources. In addition, transitions at the boundaries between surveys with different sky coverage can introduce slight morphological discontinuities in the reconstructed component maps. Transitions at the boundaries between surveys with different sky coverage can also introduce slight morphological discontinuities in the reconstructed component maps. The solution further depends on the adopted priors within \texttt{Commander}, and in low S/N regions posterior distributions are partially prior-driven. The model provides a self-consistent large-scale description of the sky, but its parameters should be interpreted with these data, model, calibration, and prior dependencies in mind.

\section{CONCLUSION} \label{sec:conclusions}
We have presented a new all-sky model of low-frequency diffuse Galactic emission describing the regime where synchrotron, free--free, and spinning dust dominate the sky brightness. This work extends the \textit{Planck} 2015 model by incorporating more recent surveys that anchor the sky closer to the CMB frequency range: S-PASS, C-BASS, and QUIJOTE (in the Galactic plane). We also used the Cosmoglobe reprocessed WMAP and \textit{Planck} LFI maps with improved large-scale calibration, replacing earlier versions of these data sets. The main advance is the re-anchoring of the synchrotron component at C-band frequencies, reducing the dependence on extrapolations from the Haslam 408\,MHz map and on the associated low-frequency curvature and large-scale survey-systematic uncertainties. The recovered spectral index distribution shows flattening toward the Galactic plane and steepening at high latitudes, consistent with variations in the cosmic-ray electron population, and leads to a redistribution of power between synchrotron, free--free, and spinning dust relative to earlier models.

The spinning dust component remains strongly correlated with thermal dust, but its total power is reduced compared to the \textit{Planck} 2015 reconstruction as a consequence of the revised synchrotron solution. The peak-frequency map exhibits coherent large-scale structure, while the spectral width remains weakly constrained outside bright regions. Free--free emission is more stable at high latitudes through the use of H$\alpha$ as a tracer, although extinction corrections and degeneracies in the Galactic plane remain sources of uncertainty. QUIJOTE data provide additional constraints in the 10--20\,GHz range in selected regions, but the dominant structural improvements arise from the revised low-frequency anchoring. A key product of this work is a new all-sky total-intensity map at 4.76\,GHz, which provides a reference tracer of diffuse synchrotron emission between classical radio surveys and microwave data. The component maps also provide updated inputs for sky simulations and foreground modeling, reducing dependence on large extrapolations from very low frequencies and improving the description of the transition between radio and microwave emission.

Residual limitations remain. Degeneracies between synchrotron, free--free, and spinning dust persist in the 10--40\,GHz regime, and absolute zero-level uncertainties, incomplete frequency coverage, and parametric assumptions affect the recovered parameters. In particular, the recovered synchrotron spectral-index map shows moderate dust-correlated structure, indicating that some residual leakage between synchrotron and spinning dust remains in the fitted spectral parameters. Compact sources with unusual spectra and beam mismatches introduce localized residuals. Addressing these issues requires future full-sky surveys in the 1--20\,GHz range with improved calibration and large-scale sensitivity. Within these limits, this work establishes an updated reference description of the low-frequency diffuse Galactic sky and defines a baseline for studies of Galactic emission, CMB foreground separation, and sky simulation, while identifying the frequency range where further observational progress is required.

\section*{Acknowledgements}
The authors thank José Alberto Rubiño-Martín and Ricardo T. Génova-Santos for useful comments. GAH acknowledges support from the Dean’s Doctoral Scholarship at the University of Manchester. CD/SEH/JPL/VS acknowledge funding from the STFC (Consolidated Grant ST/P000649/1) and CD/SEH from UKSA (LiteBIRD UK ST/Y005945/1). ACT, MEJ, and GW-D also acknowledge support from the Horizon Europe project RadioForegroundsPlus (GA 101135036), which is supported in the U.K. by UKRI grant number 10101603. This paper uses pre-publication data from the C-BASS project, which is a collaboration between Oxford and Manchester Universities in the U.K., the California Institute of Technology in the U.S., Rhodes University, UKZN and the South African Radio Astronomy Observatory in South Africa, and the King Abdulaziz City for Science and Technology (KACST) in Saudi Arabia. The work at Oxford was supported by funding from STFC, the Royal Society, and the University of Oxford. The work at the California Institute of Technology and Owens Valley Radio Observatory was supported by National Science Foundation (NSF) awards AST-0607857, AST-1010024, AST-1212217, and AST-1616227, and by NASA award NNX15AF06G. The current work has received funding from the European Union’s Horizon 2020 research and innovation programme under grant agreement numbers 819478 (ERC; \textsc{Cosmoglobe}), 101165647 (ERC, \textsc{Origins}), and 101141621 (ERC, \textsc{Commander}). This article reflects the views of the authors only. The funding body is not responsible for any use that may be made of the information contained therein. This research is also funded by the Research Council of Norway under grant agreement number 344934 (YRT; \textsc{CosmoglobeHD}). This work made use of standard scientific Python packages, including \texttt{NumPy} \citep{Harris2020}, \texttt{SciPy} \citep{Virtanen2020}, \texttt{Matplotlib} \citep{Hunter2007}, \texttt{Astropy} \citep{Astropy2013, Astropy2018, Astropy2022}, \texttt{NaMaster} \citep{Alonso2019}, and \texttt{Healpy} \citep{Zonca2019}, which is based on the HEALPix framework \citep{Gorski2005}.

\section*{Data Availability}
The input data used in this work are publicly available from the respective survey archives. \textit{Planck} HFI data were obtained from the \textit{Planck} Legacy Archive (\url{https://pla.esac.esa.int/#home}). Cosmoglobe-processed WMAP and LFI data are available from the Cosmoglobe website (\url{https://www.cosmoglobe.uio.no/}). WMAP DR5 data were obtained from the NASA LAMBDA archive (\url{https://lambda.gsfc.nasa.gov/product/}). S-PASS data are available from the S-PASS collaboration website (\url{https://sites.google.com/inaf.it/spass}). QUIJOTE data were obtained from the QUIJOTE collaboration website (\url{https://research.iac.es/proyecto/quijote/pages/en/telescopes.php}). \texttt{Commander} is open-source software and available at \url{https://github.com/Cosmoglobe/Commander}. The C-BASS data used in this analysis will be publicly available in a future data release (Taylor et al. in prep.). The derived maps presented in this work will be publicly available through the NASA LAMBDA archive upon publication.
 



\bibliographystyle{mnras}
\bibliography{refs,planck_MNRAS} 



\clearpage
\appendix
\appendix
\section{Hybrid WMAP Map Construction}\label{app:wmap_hybrid}
The Cosmoglobe WMAP maps were not beam-symmetrized during map-making, and point sources retain the elliptical shape of the native instrumental beams, most prominently in the K- and Ka-bands. In component-separation runs in pixel space, these elliptical beam patterns produce quadrupolar residuals around bright compact sources due to mismatches in source morphology across frequency channels.

\begin{figure}
    \centering
    \includegraphics[width=\linewidth]{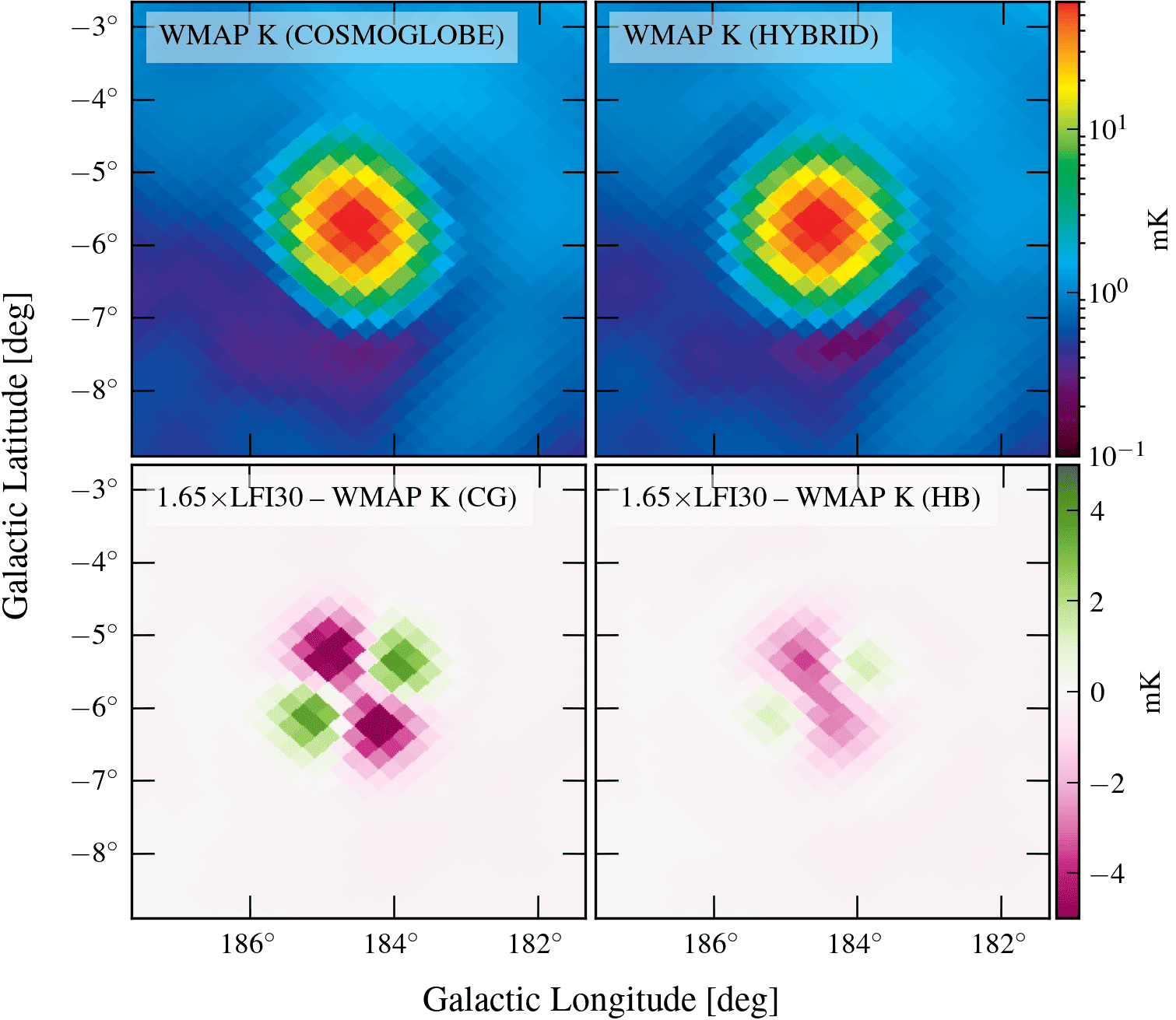}
    \caption{Effect of the hybrid map construction in the WMAP K-band. Top panels: Cosmoglobe map (left) and hybrid map (right). Bottom panels: residuals relative to the scaled \textit{Planck} LFI 30\,GHz map, highlighting the reduction of quadrupolar artifacts around bright sources in the hybrid version.}
    \label{fig:kquadrupoles}
\end{figure}

To mitigate this effect, we construct hybrid K- and Ka-band maps that combine the large-scale structure of the Cosmoglobe maps with the small-scale information of the original symmetrised WMAP DR5 maps. The combination is performed in harmonic space using two complementary window functions,
\begin{equation}
B_{\ell}^{\mathrm{L}} =
\begin{cases}
1, & \ell < \ell_0, \\[3pt]
\cos^2{\left(\dfrac{\pi}{2}\dfrac{\ell-\ell_0}{\ell_{\mathrm{f}}-\ell_0}\right)}, 
& \ell_0 \leq \ell \leq \ell_{\mathrm{f}}, \\[6pt]
0, & \ell > \ell_{\mathrm{f}},
\end{cases}
\end{equation}
and
\begin{equation}
B_{\ell}^{\mathrm{S}} =
\begin{cases}
0, & \ell < \ell_0, \\[3pt]
\cos^2{\left(\dfrac{\pi}{2}\dfrac{\ell-\ell_{\mathrm{f}}}{\ell_{\mathrm{f}}-\ell_0}\right)}, 
& \ell_0 \leq \ell \leq \ell_{\mathrm{f}}, \\[6pt]
1, & \ell > \ell_{\mathrm{f}}.
\end{cases}
\end{equation}
Here $\ell_0$ and $\ell_{\mathrm{f}}$ define the transition region in multipole space. With this definition, the Cosmoglobe harmonic coefficients are used exclusively for $\ell<\ell_0$, the original WMAP coefficients are used exclusively for $\ell>\ell_{\mathrm{f}}$, and the two maps are smoothly combined only over the transition interval $\ell_0 \leq \ell \leq \ell_{\mathrm{f}}$. The hybrid map is then
\begin{equation}
    a_{\ell m}^{\mathrm{hybrid}} =
    B_{\ell}^{\mathrm{L}} \, a_{\ell m}^{\mathrm{Cosmoglobe}}
    +
    B_{\ell}^{\mathrm{S}} \, a_{\ell m}^{\mathrm{WMAP}} .
\end{equation}
We adopt $\ell_0 = 32$ and $\ell_{\mathrm{f}} = 64$, corresponding to angular scales around the beam size after smoothing to $1^\circ$ resolution. This choice preserves the improved large-scale calibration of Cosmoglobe while restoring circularised small-scale source morphology from the original WMAP maps. The procedure is applied only to the K- and Ka-bands, since beam asymmetries in the higher-frequency channels are sufficiently small that no correction is required. Figure~\ref{fig:kquadrupoles} illustrates the impact of this correction in the K-band, showing the reduction of quadrupolar artifacts around bright compact sources.

\section{Monopole, dipole and instrumental parameters posteriors}\label{app:md}

This appendix summarizes the procedures used to determine the monopole and dipole offsets of the low-frequency surveys employed in this work. The resulting monopole, dipole, calibration, and bandpass-shift parameters are listed in Table~\ref{tab:md_inst}, and the method used to estimate the monopole of each low-frequency channel is described below.\\

\noindent\textit{C-BASS} --- The C-BASS monopole was recalibrated through a T--T analysis against the \textit{Planck} 2015 model prediction at 4.76\,GHz. The comparison was performed at $\mathrm{N_{\textrm{SIDE}}}=64$ using the PM61 mask. The fitted intercept was $21\pm1$\,mK, approximately half of the correction implied by the ARCADE\,2-based calibration. Internal runs showed that adopting this value produced stable solutions and physically plausible synchrotron spectral indices. No statistically significant dipole correction was required for C-BASS.\\

\noindent\textit{S-PASS} --- A direct T--T calibration of S-PASS against the \textit{Planck} 2015 model was found to be unreliable due to large-scale structural differences in the southern sky. Instead, morphological continuity between S-PASS and C-BASS was enforced in their overlap region. Holding the C-BASS solution fixed, five large-scale correction parameters were fitted to S-PASS: a global amplitude/spectral index scaling, a monopole term, and three dipole components. The minimization yielded an effective spectral index of \(\beta=-3.12\) and removed visible discontinuities between the surveys. The resulting monopole and dipole values were adopted as the S-PASS offsets.\\

\noindent\textit{QUIJOTE} --- QUIJOTE monopoles were estimated through T--T correlations between preliminary runs excluding QUIJOTE and the QUIJOTE maps within the adopted Galactic-plane mask. Compact sources were masked using the PS30 mask. This procedure was applied to the 11, 13, 17, and 19\,GHz channels. The offsets for all the QUIJOTE channels were consistent within \(1\sigma\) with the \cite{rubino2023quijote}. No statistically significant dipole correction was required within the plane region used.\\

\noindent\textit{Cosmoglobe WMAP and LFI} --- Cosmoglobe-reprocessed WMAP and LFI maps exhibit offset shifts of order $\sim10$\,\si{\micro\kelvin} relative to their original releases. For these channels, the \textit{Planck} 2015 monopole and dipole normalization scheme is retained, and residual offsets are adjusted through T--T correlations at $\mathrm{N_{\textrm{SIDE}}}=64$ using the PM61 mask.

\input{tables/md_inst_params}

\section{DERIVED SPECTRAL INDEX MAPS}
\label{app:si}

Although our baseline analysis directly samples the synchrotron spectral index $\beta_\mathrm{s}$ within the parametric component-separation, effective spectral index maps can also be constructed in the traditional manner from the modeled sky emission. This allows direct comparison with spectral index products from the literature defined between specific frequency pairs.

\begin{figure*}
    \centering
    \includegraphics[width=\linewidth]{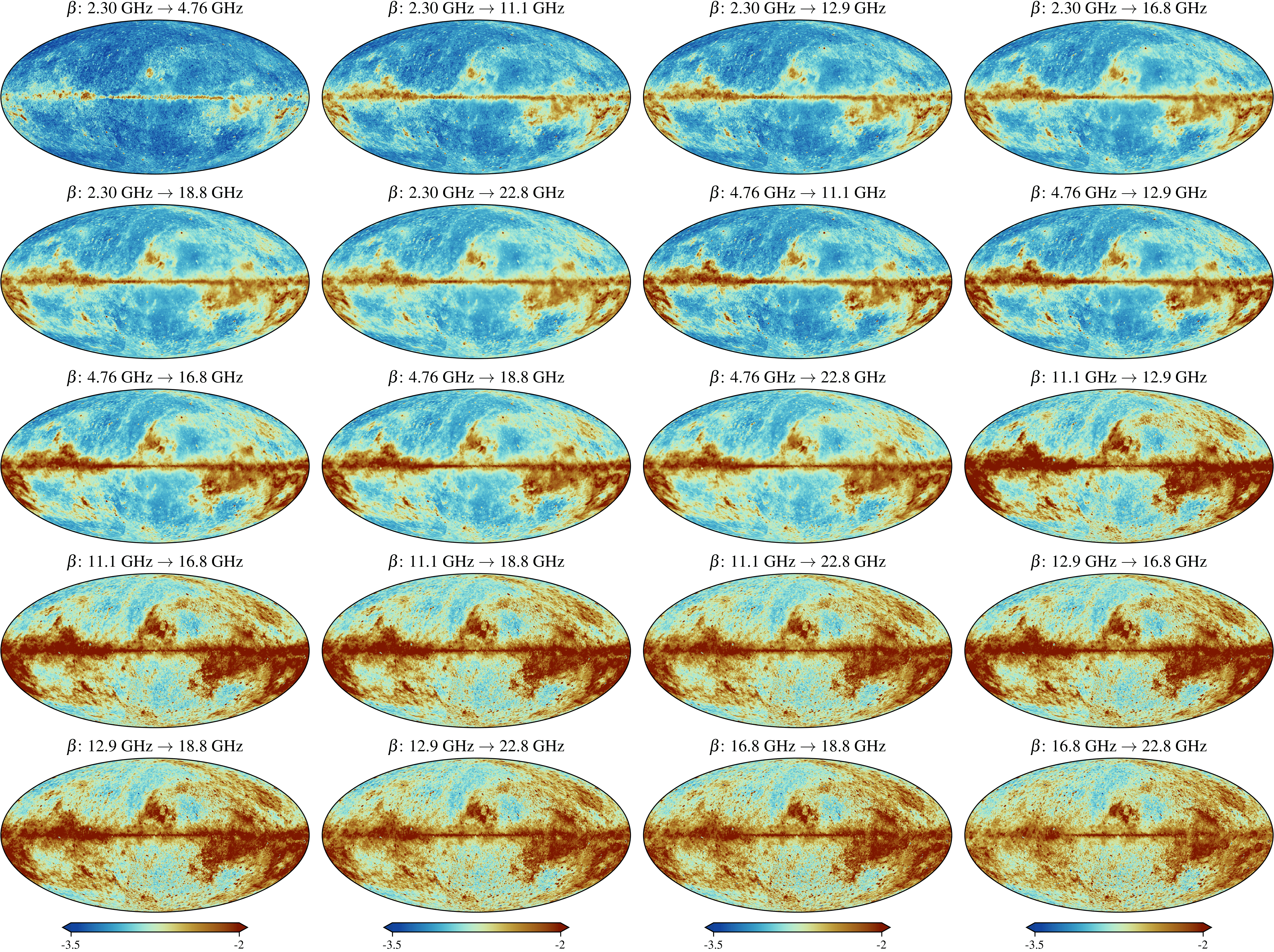}
    \caption{Examples of effective spectral index maps derived from the sky model presented in this work using different frequency pairs. The indices are computed pixel by pixel assuming a local power-law behavior of the brightness temperature (Eq.~\ref{eq:si}).}
\label{fig:si_examples}
    \label{fig:si_maps}
\end{figure*}

For two frequencies $\nu_1$ and $\nu_2$, the effective spectral index $\beta(\nu_1,\nu_2)$ is computed pixel by pixel assuming a local power-law behavior of the brightness temperature,
\begin{equation}
T(\nu) \propto \nu^{\beta},
\end{equation}
which leads to
\begin{equation}\label{eq:si}
\beta(\nu_1,\nu_2) = 
\frac{\log\!\left[T(\nu_1)/T(\nu_2)\right]}
{\log\!\left(\nu_1/\nu_2\right)}.
\end{equation}
In our case, the brightness temperatures $T(\nu_1)$ and $T(\nu_2)$ are obtained from the full sky model evaluated at the corresponding frequencies, including all components. These maps represent the effective spectral behavior of the total emission or of a specific component, depending on the model outputs used. Figure~\ref{fig:si_examples} shows examples of spectral index maps derived from our solution using different frequency pairs. These maps illustrate how the effective spectral index captures the frequency-dependent behavior of Galactic foreground emission. Spectral indices defined between different frequency pairs probe distinct spectral regimes, resulting in systematic differences in the inferred slopes. This frequency-range dependence must be taken into account when comparing spectral index maps derived from different data combinations.



\bsp	
\label{lastpage}
\end{document}

%% file: tables/data_sets.tex
\begin{table*}
    \begin{threeparttable}
    \centering
    \renewcommand{\arraystretch}{1.2}
    \caption{Overview of data sets.}
    \begin{tabular*}{\textwidth}{@{\extracolsep{\fill}}cccccccc}
    \toprule
    \toprule
    \multirow{2}{*}{Instrument} & \multicolumn{1}{c}{Frequency} & \multirow{2}{*}{Detector label} & \multirow{2}{*}{Resolution} & Noise RMS & \multirow{2}{*}{Coverage used$^{\text{ a}}$} &\multirow{2}{*}{Units} & \multirow{2}{*}{Reference} \\
                                & [GHz]                         &                      &                             & $\sigma(1^\circ)$                   &                       & \\
    \midrule
      S-PASS$^{\text{ b}}$ \hfill\ldots\ldots\ldots\hfill      & 2.30     &          &  $8.\!\!^\prime9$      & \hspace{2.4em}606 (85.8) & $\delta < -14.\!\!^\circ6$ & \si{\micro\kelvin}$_\text{RJ}$ & \cite{Carretti2019}  \\
      C-BASS \hfill\ldots\ldots\ldots\hfill      & 4.76      &          & 47$^\prime$      & 175 & $\delta > -14.\!\!^\circ6$ & \si{\micro\kelvin}$_\text{RJ}$ & Taylor et al. (in prep.)  \\
      QUIJOTE$^{\text{ c}}$ \hfill\ldots\ldots\ldots\hfill      & 11.1     &          & 55$^\prime$      & \hspace{2.1em}649 (83) & Galactic plane & \si{\micro\kelvin}$_\text{CMB}$ & \cite{rubino2023quijote}  \\
                                                  & 12.9     & & 56$^\prime$      & \hspace{2.1em}479 (65) & (see Fig. \ref{fig:masks}) & & \\
                                                  & 16.8     & & 39$^\prime$      & \hspace{2.4em}897 (130) &  & & \\
                                                  & 18.8     & & 40$^\prime$      & \hspace{2.4em}714 (180) &  & & \\
    WMAP \hfill\ldots\ldots\ldots\hfill    & 22.8        & K        & 53$^\prime$       & 6.0 & all-sky & \si{\micro\kelvin}$_\text{CMB}$ & \cite{Bennett2013} \\
                                                 & 33.0        & Ka       & 40$^\prime$       & 4.6 & &   & \cite{Watts2023}\\
                                                 & 40.6        & Q1       & 31$^\prime$       & 5.6 &   &   &  \\
                                                 &           & Q2       & 31$^\prime$       & 5.2 &   &   &  \\
                                                 & 60.8        & V1       & 21$^\prime$       & 7.6 &   &   &  \\
                                                 &           & V2       & 21$^\prime$       & 6.5 &   &   &  \\
                                                 & 93.5        & W1       & 13$^\prime$       & 10.1 &   &   &  \\
                                                 &           & W2       & 13$^\prime$       & 11.2 &   &   &  \\
                                                 &           & W3       & 13$^\prime$       & 12.1 &   &   &  \\
                                                 &           & W4       & 13$^\prime$       & 11.2 &   &   &  \\
      \textit{Planck} LFI \hfill\ldots\ldots\ldots\hfill  & 28.4        & all    & 32.\!$^\prime$4     & 2.3 & all-sky & \si{\micro\kelvin}$_\text{CMB}$ & \cite{planck2016-l02} \\
                                                 & 44.1        & all    & 27.\!$^\prime$1   & 2.7 &   &   & \cite{Watts2023}\\
                                                 & 70.3        & all    & 13.\!$^\prime$6   & 2.1 &   &   & \\
      \textit{Planck} HFI \hfill\ldots\ldots\ldots\hfill  & 100       & ds1      & 9.\!$^\prime$7      & 0.9 & all sky & \si{\micro\kelvin}$_\text{CMB}$ & \cite{planck2014-a09} \\
                                                 &           & ds2      & 9.\!$^\prime$7      & 0.8 &   &   & \\
                                                 & 143       & ds1      & 7.\!$^\prime$2      & 0.7 &   &   & \\
                                                 &           & ds2      & 7.\!$^\prime$2      & 0.7 &   &   & \\
                                                 &           & 5        & 7.\!$^\prime$2      & 0.9 &   &   & \\
                                                 &           & 6        & 7.\!$^\prime$2      & 1.1 &   &   & \\
                                                 &           & 7        & 7.\!$^\prime$2      & 1.0 &   &   & \\
                                                 & 217       & 1        & 5.\!$^\prime$0      & 1.8 &   &   & \\
                                                 &           & 2        & 5.\!$^\prime$0      & 1.9 &   &   & \\
                                                 &           & 3        & 5.\!$^\prime$0      & 1.7 &   &   & \\
                                                 &           & 4        & 5.\!$^\prime$0      & 1.8 &   &   & \\
                                                 & 353       & ds2      & 4.\!$^\prime$9      & 4.5 &   &   & \\
                                                 &           & 1        & 4.\!$^\prime$9      & 3.5 &   &   & \\
                                                 & 545       & 2        & 4.\!$^\prime$7      & 0.01 & all-sky & MJy/sr              & \\
                                                 &           & 4        & 4.\!$^\prime$7      & 0.01 &        &                & \\
                                                 & 857       & 2        & 4.\!$^\prime$4      & 0.01 &         &               & \\
    \bottomrule
    \end{tabular*}
    \begin{tablenotes}
    \item[]\hspace{-0.2em}\textbf{Notes:}\,$^{\text{(a)}}$ For S-PASS and QUIJOTE, we used a limited coverage compared to the survey's full coverage.\,$^{\text{(b)}}$ 600\,\si{\micro\kelvin} of regularization noise is added to the thermal noise RMS.\,$^{\text{(c)}}$ 5\% calibration uncertainty is added as uncorrelated noise to the thermal noise RMS.\,$^{\text{(b,c)}}$ The corresponding pure-noise RMS is given in brackets.
    \end{tablenotes}
    \label{tab:data}
    \end{threeparttable}
\end{table*}

%% file: tables/sed.tex
\begin{table*}
    \begin{threeparttable}
    \centering
    \renewcommand{\arraystretch}{1.2}
    \caption{Summary of parametric signal models.}
    \begin{tabular*}{\textwidth}{@{\extracolsep{\fill}}crclrcll}
    \toprule
    \toprule
    Component & \multicolumn{3}{c}{Free parameters and priors} & \multicolumn{3}{c}{Brightness temperature, $T_{\text{b}}\,$[\si{\micro\kelvin}$_{\text{RJ}}$]}  & Additional information \\
    \midrule
       & & & & $x$&$=$&$\frac{h\nu}{k_{\text{B}}T_{\text{cmb}}}$ & \\
       CMB \hfill\ldots\ldots\ldots\hfill & $A_{\text{cmb}}$&$\sim$&$\text{Uniform}(-\infty,\infty)$ & $g(\nu)$&$=$&$[\exp(x)-1]^2/[x^2\exp(x)]$ & $T_{\text{cmb}}=2.7255$\,K \\
       & & & & $T_{\text{b,cmb}}$&$=$&$A_{\text{cmb}}/g(\nu)$ & \\
       & & & \\
       \multirow{2}{*}{Synchrotron \hfill\ldots\hfill} & $A_{\text{s}}$&$>$&$0$ & \multirow{2}{*}{$T_{\text{b,s}}$}&\multirow{2}{*}{$=$}&\multirow{2}{*}{$A_{\text{s}}\left(\dfrac{\nu}{\nu_{0,\text{s}}}\right)^{\beta_{\text{s}}}$} & \multirow{2}{*}{$\nu_{0,\text{s}}=4.76$\,GHz}\\
       & $\beta_{\text{s}}$&$\sim$&$ N(-3.1,0.3)$ & & \\
       & & & \\
       \multirow{4}{*}{Free--free \hfill\ldots\ldots\hfill} & \multirow{4}{*}{$\ln{\text{EM}}$}&\multirow{4}{*}{$\sim$}&\multirow{4}{*}{$\text{Uniform}(-\infty,\infty)$} & \multirow{2}{*}{$g_{\text{ff}}$}&\multirow{2}{*}{$=$}&\multirow{2}{*}{$\ln{\left\{\exp{\left[5.960-\sqrt{3}/\pi\log{\left(\nu_9\,T_4^{-3/2}\right)}\right]}+e\right\}}$} & \multirow{2}{*}{$T_{\text{e}}=$ \textit{Planck} \citeyear{planck2014-a12} map}\\
       & & & & \multirow{2}{*}{$\tau$}&\multirow{2}{*}{$=$}&\multirow{2}{*}{$0.05468\,T_{\text{e}}^{-3/2}\,\nu_{9}^{-2}\,\text{EM}\,g_{\text{ff}}$} & \multirow{2}{*}{$T_4=T_{\text{e}}/10^4$}\\
       & & & & \multirow{2}{*}{$T_{\text{b,ff}}$}&\multirow{2}{*}{$=$} & \multirow{2}{*}{$T_{\text{e}}(1-e^{-\tau})\,10^6$}  & \multirow{2}{*}{$\nu_9=\nu/(10^9\,\text{Hz})$}\\
       & & & \\
       \multirow{3}{*}{Spinning dust \hfill\ldots\hfill} & $A_{\text{sd}}$&$>$&$0$ & \multirow{3}{*}{$T_{\text{b,sd}}$}&\multirow{3}{*}{$=$}&\multirow{3}{*}{$A_{\text{sd}}\,\exp{\left\{-\dfrac{1}{2}\left[\dfrac{\ln{(\nu/\nu_{\text{p}})}}{w}\right]^2\right\}}\left(\dfrac{\nu_{0,\text{sd}}}{\nu}\right)^2$} & \multirow{3}{*}{$\nu_{0,\text{sd}}=22.8$\,GHz}\\
       & $\nu_{\text{p}}$&$\sim$&$ N(25,5\,\text{GHz})$ & & \\
       & $w$&$\sim$&$ N(0.6,0.1)$ & & \\
       & & & \\
       \multirow{4}{*}{Thermal dust \hfill\ldots\hfill} & \multirow{2}{*}{$A_{\text{d}}$}&\multirow{2}{*}{$>$}&\multirow{2}{*}{$0$} & \multirow{3}{*}{$\gamma$}&\multirow{3}{*}{$=$}&\multirow{3}{*}{$\frac{h}{k_{\text{B}}T_{\text{d}}}$} & \multirow{4}{*}{$\nu_{0,\text{d}}=545$\,GHz}\\
       & \multirow{2}{*}{$\beta_{\text{d}}$}&\multirow{2}{*}{$\sim$}&\multirow{2}{*}{$ N(1.55,0.1)$} &\multirow{3}{*}{$T_{\text{b,d}}$}&\multirow{3}{*}{$=$}&\multirow{3}{*}{$A_{\text{d}}\,\left(\frac{\nu}{\nu_{0,\text{d}}}\right)^{\beta_{\text{d}}+1}\frac{\exp{(\gamma\,\nu_{0,\text{d}})-1}}{\exp{(\gamma\,\nu)-1}}$} & \\
       & \multirow{2}{*}{$T_{\text{d}}$}&\multirow{2}{*}{$\sim$}&\multirow{2}{*}{$ N(23,3\,\text{K})$} & & \\
       & & & \\
       SZ \hfill\ldots\ldots\ldots\hfill & $y_{\text{sz}}$&$>$&$0$ & $T_{\text{b,sz}}$&$=$&$10^6\,\dfrac{y_{\text{sz}}\,T_{\text{cmb}}}{g(\nu)}\left\{\dfrac{x[\exp{(x)}+1]}{\exp{(x)}-1}-4\right\}$ & \\
       & & & \\
       \multirow{7}{*}{Line emission \hfill\ldots\hfill} &\multirow{7}{*}{$A_i$} & \multirow{7}{*}{$>$} & \multirow{7}{*}{$0$}& \multirow{7}{*}{$T_{\text{b,}i}$} & \multirow{7}{*}{$=$} & 
       \multirow{7}{*}{$A_{i} h_{ij} \dfrac{F_i(\nu_j)}{F_{i}(\nu_{0,i})} \dfrac{g(\nu_{0,i})}{g(\nu_{j})}$} & 
       \multirow{4}{*}{$i\in\left\{ \begin{array}{l}
         \text{CO } J=1 \rightarrow 0 \\
         \text{CO } J=2 \rightarrow 1 \\
         \text{CO } J=3 \rightarrow 2 \\
        94/100
        \end{array} \right.$} \\
    
       & & & & & & & \\
       & & & & & & & \\
       & & & & & & & \\
       & & & & & & & $j=$ detector index \\
       & & & & & & & $h_{ij}=$ \textit{Planck} \citeyear{planck2014-a12} map\\
       & & & & & & & $F=$ unit conversion \\
    \bottomrule
    \end{tabular*}
    \begin{tablenotes}
    \item[]\hspace{-0.2em}\textbf{Note:} The symbol ``\textasciitilde'' implies that the respective parameter has a prior as given by the right-hand side distribution. 
    \end{tablenotes}
    \label{tab:sed}
    \end{threeparttable}
\end{table*}

%% file: tables/gof_stats.tex
\begin{table}
    \begin{threeparttable}
    \centering
    \renewcommand{\arraystretch}{1.2}
    \caption{Goodness-of-fit statistics.}
    \begin{tabular}{rllcc}
    \toprule
    \toprule
    && \multicolumn{2}{c}{RMS outside LM95} & Frac. res. \\
    & \vspace{-1.75em}\\
    && \multicolumn{2}{l}{--------------------------------}  & inside \\
    & Map & [\si{\micro\kelvin}]$^{\text{ a}}$ & $\sigma^{\text{res}}_\nu/\sigma^{\text{inst}}_\nu{}^{\text{ b}}$ & LM95 [\%]$^{\text{ c}}$ \\
    \midrule
      S-PASS          & 2.30 \hfill\hspace{1.em}\hfill  & 3.42 &  $0.01$    & 0.00 \\
      C-BASS          & 4.76 \hfill\hspace{1.em}\hfill   & 4.70 &  $0.03$     & 0.00 \\
      QUIJOTE         & 11.1 \hfill\hspace{1.em}\hfill   & \hspace{0.3em}\ldots &  \ldots      & 4.72 \\
                      & 12.9 \hfill\hspace{1.em}\hfill   & \hspace{0.3em}\ldots &  \ldots      & 4.53 \\
                      & 16.8 \hfill\hspace{1.em}\hfill   & \hspace{0.3em}\ldots &  \ldots      & 3.41 \\
                      & 18.8 \hfill\hspace{1.em}\hfill   & \hspace{0.3em}\ldots &  \ldots      & 4.87 \\
      WMAP            & K \hfill\hspace{1.em}\hfill         & 3.73 &  $0.62$   & 0.12 \\
                      & Ka \hfill\hspace{1.em}\hfill        & 4.59 &  $1.00$   & 0.86 \\
                      & Q1 \hfill\hspace{1.em}\hfill        & 5.33 &  $0.95$   & 0.67 \\
                      & Q2 \hfill\hspace{1.em}\hfill        & 5.10 &  $0.98$   & 0.54 \\
                      & V1 \hfill\hspace{1.em}\hfill        & 6.39 &  $0.84$   & 1.07 \\
                      & V2 \hfill\hspace{1.em}\hfill        & 5.76 &  $0.89$   & 0.92 \\
                      & W1 \hfill\hspace{1.em}\hfill        & 8.46 &  $0.84$   & 1.30 \\
                      & W2 \hfill\hspace{1.em}\hfill        & 9.99 &  $0.89$   & 1.52 \\
                      & W3 \hfill\hspace{1.em}\hfill        & \hspace{-0.5em}10.24 &  $0.85$  & 1.47 \\
                      & W4 \hfill\hspace{1.em}\hfill        & \hspace{-0.5em}10.35 &  $0.92$  & 1.51 \\
      \textit{Planck} & 28.4 \hfill\hspace{1.em}\hfill     & 1.52 &  $0.66$      & 0.16 \\
                      & 44.1 \hfill\hspace{1.em}\hfill     & 2.27 &  $0.84$      & 0.36 \\
                      & 70.3 \hfill\hspace{1.em}\hfill     & 1.96 &  $0.94$      & 0.37 \\
                      & 100-ds1 \hfill\hspace{1.em}\hfill   & 0.67 &  $0.75$   & 0.01 \\
                      & 100-ds2 \hfill\hspace{1.em}\hfill   & 0.58 &  $0.72$   & 0.01 \\
                      & 143-ds1 \hfill\hspace{1.em}\hfill   & 0.72 &  $1.02$   & 0.07 \\
                      & 143-ds2 \hfill\hspace{1.em}\hfill   & 0.68 &  $0.98$   & 0.08 \\
                      & 143-5 \hfill\hspace{1.em}\hfill     & 1.02 &  $1.14$   & 0.14 \\
                      & 143-6 \hfill\hspace{1.em}\hfill     & 1.18 &  $1.08$   & 0.10 \\
                      & 143-7 \hfill\hspace{1.em}\hfill     & 1.07 &  $1.07$   & 0.09 \\
                      & 217-1 \hfill\hspace{1.em}\hfill     & 1.71 &  $0.95$   & 0.06 \\
                      & 217-2 \hfill\hspace{1.em}\hfill     & 1.98 &  $1.04$   & 0.08 \\
                      & 217-3 \hfill\hspace{1.em}\hfill     & 1.64 &  $0.96$   & 0.07 \\
                      & 217-4 \hfill\hspace{1.em}\hfill     & 1.75 &  $0.97$   & 0.05 \\
                      & 353-ds2 \hfill\hspace{1.em}\hfill   & 3.89 &  $1.11$   & 0.02 \\
                      & 353-1 \hfill\hspace{1.em}\hfill     & 1.49 &  $0.33$   & 0.01 \\
                      & 545-2 \hfill\hspace{1.em}\hfill     & 9.32$^\text{ d}$ &  $0.93$   & 0.14 \\
                      & 545-4 \hfill\hspace{1.em}\hfill     & 9.55$^\text{ d}$ &  $0.95$   & 0.13 \\
                      & 857-2 \hfill\hspace{1.em}\hfill     & 1.24$^\text{ d}$ &  $0.12$   & 0.00 \\
    \bottomrule
    \end{tabular}
    \begin{tablenotes}
    \item[]\hspace{-0.25em}\textbf{Notes:}\,$^\text{(a)}$\,RMS residual outside the 95\% \texttt{Commander} likelihood mask for each channel except for QUIJOTE channels since their masks are about the coverage of LM95.\,$^\text{(b)}$\,Same as\,$^\text{(a)}$, but normalized with respect to the instrumental noise RMS listed in Table \ref{tab:data}.\,$^\text{(c)}$\,Median fractional residual in the complementary 5\% of the sky, covering the Galactic plane region.\,$^\text{(d)}$\,Unit is\,kJy/sr.
    \end{tablenotes}
    \label{tab:gof}
    \end{threeparttable}
\end{table}

%% file: tables/si_t-t_plots.tex
\begin{table}
\centering
\renewcommand{\arraystretch}{1.2}
\setlength{\tabcolsep}{5pt}
\caption{Spectral index T--T plot Spearman rank coefficients in the PM61 and LM95 regions. Point sources are masked with PS30 in both cases.}
\begin{threeparttable}
\begin{tabular*}{\columnwidth}{@{\extracolsep{\fill}}lcccc}
\toprule
\toprule
Reference $^\mathrm{a}$ 
& Freq. range & $r_\mathrm{s}$ (PM61) & $r_\mathrm{s}$ (LM95) \\
\midrule

\cite{Irfan2026} $^\mathrm{b}$ 
& @ 4.76~GHz & 0.04 & 0.22 \\

\cite{Irfan2023} $^\mathrm{b}$ 
& 0.408--23 GHz & 0.08 & 0.16 \\



\cite{Platania2003} $^\mathrm{c}$ 
& 0.408--2.3 GHz & 0.16 & 0.26 \\

\cite{Giardino2002} $^\mathrm{c}$ 
& 0.408--2.3 GHz &0.27& 0.24 \\

\bottomrule
\end{tabular*}
\begin{tablenotes}
\item[]\hspace{-0.2em}\textbf{Notes:}\,
$^{\text{(a)}}$ Spearman rank coefficients are computed from T--T plots in which the spectral index maps from this work are used as the \(x\)-axis quantity and the corresponding maps from external studies as the \(y\)-axis quantity.\,
$^{\text{(b)}}$ Compared against the synchrotron spectral index map at 4.76\,GHz from this work.\,
$^{\text{(c)}}$ Compared against the 2.30--4.76\,GHz spectral index map generated from the model of this work.
\end{tablenotes}
\end{threeparttable}
\label{tab:t-t}
\end{table}

%% file: tables/md_inst_params.tex
\begin{table*}
    \begin{threeparttable}
    \centering
    \renewcommand{\arraystretch}{1.2}
    \caption{Monopoles, dipoles, calibration factors and bandpass corrections derived within the baseline temperature model.}
    \begin{tabular*}{\textwidth}{@{\extracolsep{\fill}}c c c A A A A A A}
    \toprule
    \toprule
    \multirow{2}{*}{Survey}
    & Frequency
    & Detector
    & \multicolumn{3}{c}{Monopole}
    & \multicolumn{3}{c}{$X$ dipole}
    & \multicolumn{3}{c}{$Y$ dipole}
    & \multicolumn{3}{c}{$Z$ dipole}
    & \multicolumn{3}{c}{Calibration}
    & \multicolumn{3}{c}{Bandpass shift} \\
    & [GHz]
    & label
    & \multicolumn{3}{c}{[\si{\micro\kelvin}]}
    & \multicolumn{3}{c}{[\si{\micro\kelvin}]}
    & \multicolumn{3}{c}{[\si{\micro\kelvin}]}
    & \multicolumn{3}{c}{[\si{\micro\kelvin}]}
    & \multicolumn{3}{c}{[$\%$]}
    & \multicolumn{3}{c}{[GHz]} \\
    \midrule
    S-PASS \hfill\ldots\ldots\ldots\hfill
    & $2.30$ & \ldots
    & \intnum{14}{6^{\,\text{a,b}}}
    & \intnum{3}{6^{\,\text{a,b}}}
    & \intnum{6}{2^{\,\text{a,b}}}
    & \intnum{6}{5^{\,\text{a,b}}}
    & \anchornum{}{0^{\,\text{a}}}{}
    & \anchornum{}{0^{\,\text{a}}}{} \\

    C-BASS \hfill\ldots\ldots\ldots\hfill
    & $4.76$ & \ldots
    & \intnum{2}{1^{\,\text{a,b}}}
    & \anchornum{}{0^{\,\text{a}}}{}
    & \anchornum{}{0^{\,\text{a}}}{}
    & \anchornum{}{0^{\,\text{a}}}{}
    & \anchornum{}{0^{\,\text{a}}}{}
    & \anchornum{}{0^{\,\text{a}}}{} \\

    QUIJOTE \hfill\ldots\ldots\ldots\hfill
    & $11.1$ & \ldots
    & \intnum{-82}{6^{\,\text{a}}}
    & \anchornum{}{0^{\,\text{a}}}{}
    & \anchornum{}{0^{\,\text{a}}}{}
    & \anchornum{}{0^{\,\text{a}}}{}
    & \anchornum{}{0^{\,\text{a}}}{}
    & \anchornum{}{0^{\,\text{a}}}{} \\

    & $12.9$ & \ldots
    & \intnum{-49}{0^{\,\text{a}}}
    & \anchornum{}{0^{\,\text{a}}}{}
    & \anchornum{}{0^{\,\text{a}}}{}
    & \anchornum{}{0^{\,\text{a}}}{}
    & \anchornum{}{0^{\,\text{a}}}{}
    & \anchornum{}{0^{\,\text{a}}}{} \\

    & $16.8$ & \ldots
    & \intnum{-52}{9^{\,\text{a}}}
    & \anchornum{}{0^{\,\text{a}}}{}
    & \anchornum{}{0^{\,\text{a}}}{}
    & \anchornum{}{0^{\,\text{a}}}{}
    & \anchornum{}{0^{\,\text{a}}}{}
    & \anchornum{}{0^{\,\text{a}}}{} \\

    & $18.8$ & \ldots
    & \intnum{-94}{2^{\,\text{a}}}
    & \anchornum{}{0^{\,\text{a}}}{}
    & \anchornum{}{0^{\,\text{a}}}{}
    & \anchornum{}{0^{\,\text{a}}}{}
    & \anchornum{}{0^{\,\text{a}}}{}
    & \anchornum{}{0^{\,\text{a}}}{} \\

    WMAP \hfill\ldots\ldots\ldots\hfill
    & $22.8$ & K
    & \pmnum{-31}{1}
    & \pmnum{-6.5}{0.2}
    & \pmnum{5.1}{0.1}
    & \pmnum{-6.8}{0.1}
    & \anchornum{}{0^{\,\text{a}}}{}
    & \anchornum{}{0^{\,\text{a}}}{} \\

    & $33.0$ & Ka
    & \intnum{-1}{5^{\,\text{a}}}
    & \pmnum{-3.8}{0.1}
    & \pmnum{4.6}{0.1}
    & \pmnum{-7.9}{0.1}
    & \pmnum{-0.0}{0.1}
    & \anchornum{}{0^{\,\text{a}}}{} \\

    & $40.6$ & Q1
    & \pmnum{-10}{1}
    & \pmnum{-3.0}{0.1}
    & \pmnum{3.5}{0.1}
    & \pmnum{-6.3}{0.1}
    & \pmnum{-0.1}{0.1}
    & \anchornum{}{0^{\,\text{a}}}{} \\

    &  & Q2
    & \pmnum{-9}{1}
    & \pmnum{-3.1}{0.1}
    & \pmnum{3.7}{0.1}
    & \pmnum{-6.4}{0.1}
    & \pmnum{0.1}{0.1}
    & \pmnum{0.2}{0.1} \\

    & $60.8$ & V1
    & \pmnum{-3}{1}
    & \pmnum{-2.7}{0.1}
    & \pmnum{1.2}{0.1}
    & \pmnum{-3.7}{0.1}
    & \pmnum{0.1}{0.1}
    & \anchornum{}{0^{\,\text{a}}}{} \\

    &  & V2
    & \pmnum{-3}{1}
    & \pmnum{-2.7}{0.1}
    & \pmnum{1.2}{0.1}
    & \pmnum{-3.6}{0.1}
    & \pmnum{0.4}{0.1}
    & \pmnum{-0.2}{0.1} \\

    & $93.5$ & W1
    & \pmnum{-15}{1}
    & \pmnum{-1.8}{0.1}
    & \pmnum{-0.6}{0.1}
    & \pmnum{-2.9}{0.1}
    & \pmnum{0.4}{0.1}
    & \anchornum{}{0^{\,\text{a}}}{} \\

    &  & W2
    & \pmnum{-13}{1}
    & \pmnum{-2.1}{0.1}
    & \pmnum{-0.5}{0.1}
    & \pmnum{-2.7}{0.1}
    & \pmnum{0.6}{0.1}
    & \pmnum{-0.2}{0.4} \\

    &  & W3
    & \pmnum{-17}{1}
    & \pmnum{-1.8}{0.1}
    & \pmnum{-1.0}{0.1}
    & \pmnum{-3.0}{0.1}
    & \pmnum{-0.0}{0.1}
    & \pmnum{0.9}{0.5} \\

    &  & W4
    & \pmnum{-14}{1}
    & \pmnum{-1.9}{0.1}
    & \pmnum{-0.5}{0.1}
    & \pmnum{-2.6}{0.1}
    & \pmnum{0.4}{0.1}
    & \pmnum{-0.1}{0.4} \\

    \textit{Planck} LFI \hfill\ldots\ldots\ldots\hfill
    & $28.4$ & all
    & \pmnum{-20}{1}
    & \intnum{-}{6^{\,\text{a}}}
    & \intnum{}{3^{\,\text{a}}}
    & \intnum{-}{6^{\,\text{a}}}
    & \pmnum{0.6}{0.1}
    & \pmnum{0.3}{0.1} \\

    & $44.1$ & all
    & \pmnum{-8}{1}
    & \pmnum{-4.4}{0.1}
    & \pmnum{0.5}{0.1}
    & \pmnum{-3.4}{0.1}
    & \pmnum{0.2}{0.1}
    & \pmnum{0.0}{0.1} \\

    & $70.3$ & all
    & \pmnum{-3}{1}
    & \pmnum{-3.8}{0.1}
    & \pmnum{-1.9}{0.1}
    & \pmnum{-0.6}{0.1}
    & \pmnum{0.8}{0.1}
    & \pmnum{0.0}{0.1} \\

    \textit{Planck} HFI \hfill\ldots\ldots\ldots\hfill
    & $100$ & ds1
    & \intnum{}{9^{\,\text{a}}}
    & \anchornum{}{0^{\,\text{a}}}{}
    & \anchornum{}{0^{\,\text{a}}}{}
    & \anchornum{}{0^{\,\text{a}}}{}
    & \pmnum{0.11}{0.01}
    & \pmnum{0.4}{0.6} \\

    &  & ds2
    & \pmnum{7}{1}
    & \pmnum{0.1}{0.1}
    & \pmnum{0.0}{0.1}
    & \pmnum{0.1}{0.1}
    & \pmnum{0.07}{0.01}
    & \pmnum{0.6}{0.6} \\

    & $143$ & ds1
    & \intnum{2}{1^{\,\text{a}}}
    & \anchornum{}{0^{\,\text{a}}}{}
    & \anchornum{}{0^{\,\text{a}}}{}
    & \anchornum{}{0^{\,\text{a}}}{}
    & \anchornum{}{0^{\,\text{a}}}{}
    & \pmnum{0.9}{0.1} \\

    &  & ds2
    & \pmnum{22}{1}
    & \pmnum{0.1}{0.1}
    & \pmnum{0.0}{0.1}
    & \pmnum{-0.1}{0.1}
    & \pmnum{-0.04}{0.01}
    & \pmnum{0.0}{0.1} \\

    &  & 5
    & \pmnum{21}{1}
    & \pmnum{-0.5}{0.1}
    & \pmnum{0.0}{0.1}
    & \pmnum{-0.2}{0.1}
    & \pmnum{0.09}{0.01}
    & \pmnum{-0.2}{0.1} \\

    &  & 6
    & \pmnum{21}{1}
    & \pmnum{-0.4}{0.1}
    & \pmnum{0.0}{0.1}
    & \pmnum{-0.2}{0.1}
    & \pmnum{0.12}{0.01}
    & \pmnum{0.5}{0.1} \\

    &  & 7
    & \pmnum{21}{1}
    & \pmnum{-0.2}{0.1}
    & \pmnum{0.0}{0.1}
    & \pmnum{-0.1}{0.1}
    & \pmnum{0.01}{0.01}
    & \pmnum{-0.1}{0.1} \\

    & $217$ & 1
    & \pmnum{63}{1}
    & \pmnum{-1.3}{0.1}
    & \pmnum{-3.5}{0.1}
    & \pmnum{3.9}{0.1}
    & \anchornum{}{0^{\,\text{a}}}{}
    & \pmnum{-0.2}{0.1} \\

    &  & 2
    & \pmnum{63}{1}
    & \pmnum{-1.2}{0.1}
    & \pmnum{-3.5}{0.1}
    & \pmnum{4.0}{0.1}
    & \pmnum{0.01}{0.01}
    & \pmnum{-0.3}{0.1} \\

    &  & 3
    & \pmnum{61}{1}
    & \pmnum{-1.5}{0.1}
    & \pmnum{-3.7}{0.1}
    & \pmnum{4.1}{0.1}
    & \pmnum{-0.03}{0.01}
    & \pmnum{-0.1}{0.1} \\

    &  & 4
    & \pmnum{63}{1}
    & \pmnum{-1.0}{0.1}
    & \pmnum{-3.7}{0.1}
    & \pmnum{4.1}{0.1}
    & \pmnum{0.02}{0.01}
    & \pmnum{-0.3}{0.1} \\

    & $353$ & ds2
    & \pmnum{428}{2}
    & \pmnum{-4}{1}
    & \pmnum{-7}{1}
    & \pmnum{7}{1}
    & \pmnum{-0.0}{0.1}
    & \pmnum{0.2}{0.1} \\

    &  & 1
    & \pmnum{427}{3}
    & \pmnum{-5}{1}
    & \pmnum{-17}{1}
    & \pmnum{17}{1}
    & \pmnum{0.3}{0.1}
    & \pmnum{0.0}{0.1} \\

    & $545$ & 2
    & \decnum{0}{37^{\,\text{a,c}}}
    & \anchornum{}{0^{\,\text{a}}}{}
    & \anchornum{}{0^{\,\text{a}}}{}
    & \anchornum{}{0^{\,\text{a}}}{}
    & \decnum{-2}{8^{\,\text{a}}}
    & \decnum{2}{0^{\,\text{a}}} \\

    &  & 4
    & \pmnum{0.36}{0.01^{\,\text{c}}}
    & \anchornum{}{0^{\,\text{a}}}{}
    & \anchornum{}{0^{\,\text{a}}}{}
    & \anchornum{}{0^{\,\text{a}}}{}
    & \decnum{-3}{2^{\,\text{a}}}
    & \decnum{2}{8^{\,\text{a}}} \\

    & $857$ & 2
    & \pmnum{0.64}{0.01^{\,\text{c}}}
    & \anchornum{}{0^{\,\text{a}}}{}
    & \anchornum{}{0^{\,\text{a}}}{}
    & \anchornum{}{0^{\,\text{a}}}{}
    & \decnum{1}{7^{\,\text{a}}}
    & \decnum{5}{8^{\,\text{a}}} \\

    \bottomrule
    \end{tabular*}
    \begin{tablenotes}
    \item[]\textbf{Notes:}\,$^{\text{(a)}}$ Fixed at reference value.\,$^{\text{(b)}}$ Unit is mK.\,$^{\text{(c)}}$ Unit is MJy/sr.
    \end{tablenotes}
    \label{tab:md_inst}
    \end{threeparttable}
\end{table*}